\pdfoutput=1
\documentclass[10pt]{article}
\usepackage[colorlinks=true,linkcolor=black,citecolor=black,urlcolor=black,bookmarks=true,breaklinks=true]{hyperref}
\usepackage{float}
\usepackage[symbol]{footmisc}
\usepackage[superscript,sort]{cite}
\usepackage{array}
\usepackage{bm}
\usepackage{longtable}

\newcolumntype{x}[1]{%
	>{\centering\hspace{0pt}}p{#1}}%
\usepackage[normalem]{ulem} 
\usepackage{amsmath,amssymb}

\usepackage{amsthm,amscd,amsxtra,amsfonts,amsmath,amssymb,multirow}
\usepackage{wrapfig}
\usepackage[tiny,compact]{titlesec}
\usepackage{threeparttable} 
\usepackage{helvet}

\usepackage{booktabs}
\usepackage[table]{xcolor}
\usepackage{xcolor,colortbl}
\usepackage{placeins}
\usepackage{tikz}
\usetikzlibrary{shapes,arrows}
\usepackage[T1]{fontenc}
\usepackage[linesnumbered,ruled]{algorithm2e} 
\titlespacing*{\section}{0pt}{*0}{*0}
\titlespacing*{\subsection}{0pt}{*0}{*0}
\titlespacing*{\subsubsection}{0pt}{*0}{*0}
\titlespacing{\paragraph}{0pt}{*0}{*1}

\definecolor{MyPurple}{rgb}{1,0,1}

\newcommand{\beq}[1]{\begin{equation} \label{#1}}
\newcommand{\eeq}{\end{equation}}
\newcommand{\barray}{\begin{array}{ll}}
	\newcommand{\earray}{\end{array}}

\usepackage{longtable} 
\usepackage{threeparttable} 

\usepackage{array}
\newcolumntype{P}[1]{>{\centering\arraybackslash}p{#1}}

\usepackage{multirow}
\usepackage{color, colortbl}

\definecolor{Lightblue}{rgb}{0.867,0.914,0.961}
\definecolor{Lightgreen}{rgb}{0.883,0.934,0.848}

\DeclareGraphicsExtensions{.png}

\title{Are 2D  fingerprints still valuable for drug discovery?}
\author{Kaifu Gao$^{1}$, Duc Duy Nguyen$^{1}$, 
Vishnu Sresht$^{2}$,
Alan M. Mathiowetz$^{2}$,
Meihua Tu$^{2}$, and  Guo-Wei Wei$^{1,3,4,}$\footnote{
Corresponding to Guo-Wei Wei.		Email: wei@math.msu.edu}\\
$^1$ Department of Mathematics,
Michigan State University, MI 48824, USA.\\
$^2$ Pfizer Medicine Design, 610 Main St, Cambridge, MA 02139, USA.\\
$^3$ Department of Electrical and Computer Engineering,
Michigan State University, MI 48824, USA. \\
$^4$ Department of Biochemistry and Molecular Biology,
Michigan State University, MI 48824, USA. \\
}

\date{\today}

\begin{document}

\maketitle

\begin{abstract}
 Recently, molecular fingerprints extracted from three-dimensional (3D) structures using advanced mathematics, such as algebraic topology, differential geometry, and graph theory have been paired with efficient machine learning, especially deep learning algorithms to outperform other methods in drug discovery applications and competitions. This raises the question of whether classical 2D fingerprints are still valuable in computer-aided drug discovery.  This work considers 23 datasets associated with four typical problems, namely protein-ligand binding, toxicity, solubility and partition coefficient to assess the performance of eight 2D fingerprints. Advanced machine learning algorithms including random forest, gradient boosted decision tree, single-task deep neural network and multitask deep neural network are employed to construct efficient 2D-fingerprint based models. Additionally, appropriate consensus models are built to further enhance the performance of 2D-fingerprint-based methods. It is demonstrated that 2D-fingerprint-based models perform as well as the state-of-the-art 3D structure-based models for the predictions of toxicity, solubility, partition coefficient and protein-ligand binding affinity based on only ligand information. However, 3D structure-based models outperform 2D fingerprint-based methods in complex-based protein-ligand binding affinity predictions. 

\end{abstract}

\section{Introduction}
 Drug discovery is  a  multi-parameter  optimization  process,  which  involves a long list of chemical, biological, and physiological  properties \cite{di2015drug}. For a drug candidate, numerous drug-related properties must be assessed, including binding affinity, toxicity, octanol-water partition coefficient (Log P), aqueous solubility (Log S), etc. Binding affinity assesses the strength of a drug's binding to its target \cite{henriksen2015computational,gao2015binding}, while, toxicity is a measure of the degree to which a chemical compound can damage an organism adversely \cite{wu2018quantitative}. In addition, a partition coefficient is defined as the ratio of concentrations of a solute in a mixture of two immiscible solvents at equilibrium and, in the case of log P, represents the drug-relatedness of a compound as well as its hydrophobic effect on human bodies  \cite{wu2018topp}. Another relevant drug attribute is aqueous solubility which plays a vital role in distribution, absorption, and biological activity, among other processes because 65-90 \%  of body mass is water \cite{lipinski1997experimental,di2006biological}. Their importance to drug design and discovery has been emphasized by many recent surveys \cite{hopkins2014role,atallah2013admet}. Indeed, unsatisfactory toxicity or pharmacokinetic properties  are responsible for approximately half of drug candidate failures to reach the market \cite{van2003admet}.

\paragraph{}
Traditional experiments for measuring drug properties are conducted either in vivo or in vitro. Such experiments are quite time consuming and expensive. Additionally,  testing with animals can raise important ethical concerns. Therefore, various computer-aided or in silico methods become more attractive since they can produce quick results without sacrificing much accuracy in many situations. Among them, one of the most popular approaches is the quantitative structure-activity/property relationship (QSAR/QSPR) analysis. It assumes that similar molecules have similar bioactivities or physicochemical properties \cite{myint2012molecular}. Based on this assumption, activities and properties of new molecules can be predicted by studying the correlation between chemical or structural features of molecules and their activities or properties, reducing the need for time-consuming experiments.

\paragraph{}
Molecular fingerprints are one way of encoding the structural features of a molecule. They play a fundamental role in QSAR/QSPR analysis, virtual screening, similarity-based compound search, target molecule ranking, drug ADMET prediction, and other drug discovery processes. Molecular fingerprints are property profiles of a molecule, usually in the form of  vectors with each vector element indicating the existence, the degree or the frequency of one particular structure feature \cite{geppert2010current,roy2012electrotopological,tareq2010predictions}. Various fingerprints have been developed for molecular feature encoding in the past few decades \cite{rogers2010extended,lo2018machine,cereto2015molecular}. Most fingerprints are 2D fingerprints which can be extracted from molecular connection tables without 3D structure information.  However, high dimensional fingerprints have also been developed to utilize  3D molecular structure and other information \cite{verma20103d}. 

\paragraph{}
There are four main categories of 2D fingerprints, namely substructure key-based fingerprints, topological or path-based fingerprints, circular fingerprints, and pharmacophore fingerprints. Substructure key-based fingerprints are bit strings representing the presence of certain substructures or fragments from a given list of structural keys in the compound. Molecular access system (MACCS) \cite{durant2002reoptimization} is one of the most popular substructure key-based fingerprint methods. Topological or path-based fingerprints are based on analyzing all the fragments of a molecule following a (usually linear) path up to a certain number of bonds, and then hashing every one of these paths to create one fingerprint. The most prominent ones in this category are FP2 \cite{o2011open}, Daylight \cite{daylight} and electro-topological state (Estate) \cite{hall1995electrotopological} fingerprints. Circular fingerprints are also hashed topological fingerprints but rather than looking for paths in a molecule, they record the environment of each atom up to a pre-determined radius. A well-known example for this class is extended-connectivity fingerprint (ECFP) \cite{rogers2010extended}. Pharmacophore fingerprints include the relevant features and interactions needed for a molecule to be active against a given target, including  2D-pharmacophore \cite{landrum2006rdkit}, 3D-pharmacophore \cite{landrum2006rdkit} and extended reduced graph (ERG) \cite{stiefl2006erg} fingerprints as examples. Since 2D fingerprints only rely on the 2D structures, their generation is easy, fast and convenient.

\paragraph{}
In addition to the four categories mentioned above, recent improvements in deep learning have enabled the creation of neural fingerprints \cite{duvenaud2015convolutional,yang2019learned}— where the mapping between fingerprints and 2D structures is learned simultaneously with the parameters of the regression/classification model that maps fingerprints to targets. These ‘learned’ fingerprints can potentially improve predictive performance on QSAR/QSPR tasks, but they must be relearned when trying to predict new properties across significantly different regions of chemical space. Since the focus of this work is on comparing 2D and 3D descriptors across a number of disparate tasks and chemically diverse datasets, we have chosen not to consider neural fingerprints.

\paragraph{}
Most commonly used 2D molecular fingerprints were derived over a decade ago and their validation was carried out using classical regression or classification algorithms, such as linear regression, logistic regression, logistic classification, naive Bayes, k-nearest neighbors,  support vector machine, etc. On the other hand, new 3D structure-based fingerprints built from  algebraic topology \cite{cang2018integration,cang2018representability}, differential geometry \cite{nguyen2019dg}, geometric graph theory \cite{nguyen2017rigidity, bramer2018multiscale}, and algebraic graph theory \cite{nguyen2019mathematical} have been developed in recent years. In particular, these new fingerprints were mostly paired with advanced machine learning algorithms, such as  random forest (RF) \cite{svetnik2003random}, gradient boosting decision tree (GBDT) \cite{schapire2003boosting}, single-task deep neural networks (ST-DNNs) \cite{basheer2000artificial},   multi-task deep neural networks  (MT-DNNs) \cite{caruana1997multitask},  convolutional neural network (CNN),  recurrent neural network (RNN), etc. methodology, which are now easily accessible to the scientific community via user-friendly deep learning frameworks in popular programming languages \cite{tensorflow2015-whitepaper, paszke2017automatic}. Often, these new methods have demonstrated higher accuracy or better performance than earlier methods in the literature, which are typically based on 2D fingerprints and/or simple machine learning algorithms for drug discovery related applications, such as protein-ligand binding  \cite{cang2018integration},  virtual screening  \cite{cang2018representability}, toxicity \cite{wu2018quantitative},  solubility \cite{ wu2018topp}, partition coefficient \cite{wu2018topp},  as well as protein folding stability change upon mutation \cite{cang2017analysis}.   Additionally, recent results from D3R Grand Challenges, a community-wide annual competition series in computer-aided drug design, indicate that structure-based methods using sophisticated 3D structure-based fingerprints have an advantage over  ligand-based methods using 2D fingerprints in scoring and free energy predictions \cite{gaieb2019d3r,nguyen2019mathematical}. These developments raise an interesting question of whether 2D fingerprints are still valuable for drug design and discovery.  Therefore, there is pressing need to reassess 2D fingerprints with advanced machine learning algorithms and compare their performance with the state-of-the-art 3D structure-based fingerprints for drug discovery related applications. 

\paragraph{}
The objective of the present work is to reassess the predictive power of eight popular 2D fingerprints for four important drug-related problems, namely, toxicity, binding affinity, Log P, and Log S, involving a total of 23 datasets. These problems are selected for the availability of reference results generated by the state-of-the-art 3D structure-based fingerprints in the literature.  To optimize 2D fingerprints' performance, advanced machine learning algorithms, including RF, GBDT, ST-DNN, and MT-DNN, are employed in the present study. Additionally,  consensus  models are  constructed from appropriate combinations of 2D fingerprint-based predictions to further enhance their performance. The predictive power of each 2D fingerprint for certain functional groups is analyzed. Extensive numerical studies  over 23 datasets using eight 2D fingerprints and four different machine learning algorithms indicate that the combination of appropriate machine learning algorithms and 2D fingerprint-based models, particularly consensus models, can bring significant improvements over previous 2D QSPR approaches especially on toxicity predictions \cite{martin2016user}. Moreover, 2D fingerprint-based models perform as well as the state-of-the-art 3D structure-based fingerprints in the predictions of toxicity, solubility, partition coefficient and ligand-based protein-ligand binding affinity. Finally,  topology-based fingerprints extracted from 3D protein-ligand complexes  have a significant advantage over 2D fingerprints in complex-based protein-ligand binding affinity predictions.  We believe that the present performance analysis and assessment will provide a useful guideline on how to choose appropriate  fingerprints and machine learning methods for drug discovery related applications.

\section{Methods}

\subsection{2D fingerprints}

\paragraph{}

In the present work, we investigate eight popular 2D  fingerprints, including FP2 fingerprint, MACCS fingerprint, Daylight fingerprint, Estate1 fingerprint, Estate2 fingerprint, ECFP4 Fingerprint, 2D-pharmacophore (Pharm2D), and extended reduced graph fingerprint (ERG). They are chosen to represent four main 2D molecular fingerprint categories, namely key-based fingerprints, topological or path-based fingerprints, circular fingerprints, pharmacophore fingerprints. These features are some of the most popular and commonly used ones.
Table \ref{table:fg} summarizes the information related to these fingerprints.
All 2D fingerprints were generated by Openbabel (version 2.4.1) \cite{o2011open} and RDKit (version 2018.09.3) \cite{landrum2006rdkit}.

\begin{table}[h]
	\centering
	\begin{tabular}{|c|P{6cm}|c|c|}
		\rowcolor{Lightgreen}
		\hline
		\textbf{Fingerprint} & \textbf{Description} & \textbf{Number of features} & \textbf{Package} \\
		\hline
		FP2 & \multicolumn{1}{m{6cm}|}{A path-based fingerprint which indexes small molecule fragments based on linear segments of up to 7 atoms \cite{o2011open}} & 256 & Openbabel \cite{o2011open} \\
		\hline
		Daylight & \multicolumn{1}{m{6cm}|}{A path-based fingerprint consisting 2048 bits and encoding all connectivity pathways in a given length through a molecule \cite{daylight}} & 2048 & \multirow{33}{*}{RDKit \cite{landrum2006rdkit}} \\
		\cline{1-3}
		MACCS & \multicolumn{1}{m{6cm}|}{A substructure keys-based fingerprint with 166 structural keys based on SMARTS patterns \cite{durant2002reoptimization}}  & 166 & \\
		\cline{1-3}
		Estate1 & \multicolumn{1}{m{6cm}|}{A topological fingerprint based on electro-topological State Indices, which encodes the intrinsic electronic state of the atom as perturbed by the electronic influence of all other atoms in the molecule within the context of the topological character of the molecule. Estate 1 represents the number of times each atom type is hit \cite{hall1995electrotopological}} & 79 &  \\
		\cline{1-3}
		Estate2 & \multicolumn{1}{m{6cm}|}{Similar to estate 1, however it contains the sum of the EState indices for atoms of each type \cite{hall1995electrotopological}} & 79 &  \\
		\cline{1-3}
		ECFP4 & \multicolumn{1}{m{6cm}|}{The de facto standard circular fingerprint based on the Morgan algorithm \cite{morgan1965generation}, which uses an iterative process to assign numeric identifiers to each atom \cite{rogers2010extended}} & 2048 &  \\
		\cline{1-3}
		Pharm2D & \multicolumn{1}{m{6cm}|}{Each bit corresponds to a particular combination of features and interactions needed for a molecule to be active against a given target \cite{landrum2006rdkit}} & 990 &  \\
		\cline{1-3}
		ERG & \multicolumn{1}{m{6cm}|}{A Pharmacophore fingerprint, which is an extended reduced graph approach using pharmacophore-type node descriptions to encode the relevant molecular properties \cite{stiefl2006erg}} & 315 & \\
		\hline
	\end{tabular}
	\caption{A introduction of eight fingerprints used in the present study.}
	\label{table:fg}
\end{table}

\subsection{Ensemble methods}

\paragraph{}
Two popular ensemble methods were used in our work. The first method is random forest (RF), which constructs a multitude of decision trees during a training process. RF can be used to predict a classification label (classification model) or a mean prediction (regression model) of the individual trees. It is very robust against overfitting and easy to use. 
The second method is gradient boosting decision tree (GBDT). In this approach, individual decision trees are combined in a stage-wise fashion to achieve the capability of learning complex features. It uses both gradient and boosting strategies to reduce model errors.  Compared to deep neural network (DNN) approaches, these two ensemble methods are robust, relatively insensitive to hyper parameters, and easy to implement. Moreover, they are much faster to train than DNN is. In fact, for small datasets, RF and GBDT can perform even better than DNN or other deep learning algorithms. Therefore, these methods have been applied to a variety of QSAR prediction problems, such as  toxicity, solvation, and binding affinity predictions \cite{wu2018quantitative,martin2016user,wang2018breaking,cang2018integration,wang2017feature}.

\subsection{Single-task deep neural network (ST-DNN)}
\paragraph{}
A DNN mimics the learning process of a biological brain by constructing a wide and deep architecture of numerous connected neuron units. A typical deep neural network often includes multiple hidden layers. In each layer, there are hundreds or even thousands of neurons. During learning stages, weights on each layer are updated by backpropagation. With a complex and deep network, DNN is capable of constructing hierarchical features and model complex nonlinear relationships.

\begin{figure}[h]
	\includegraphics[width=0.8\textwidth]{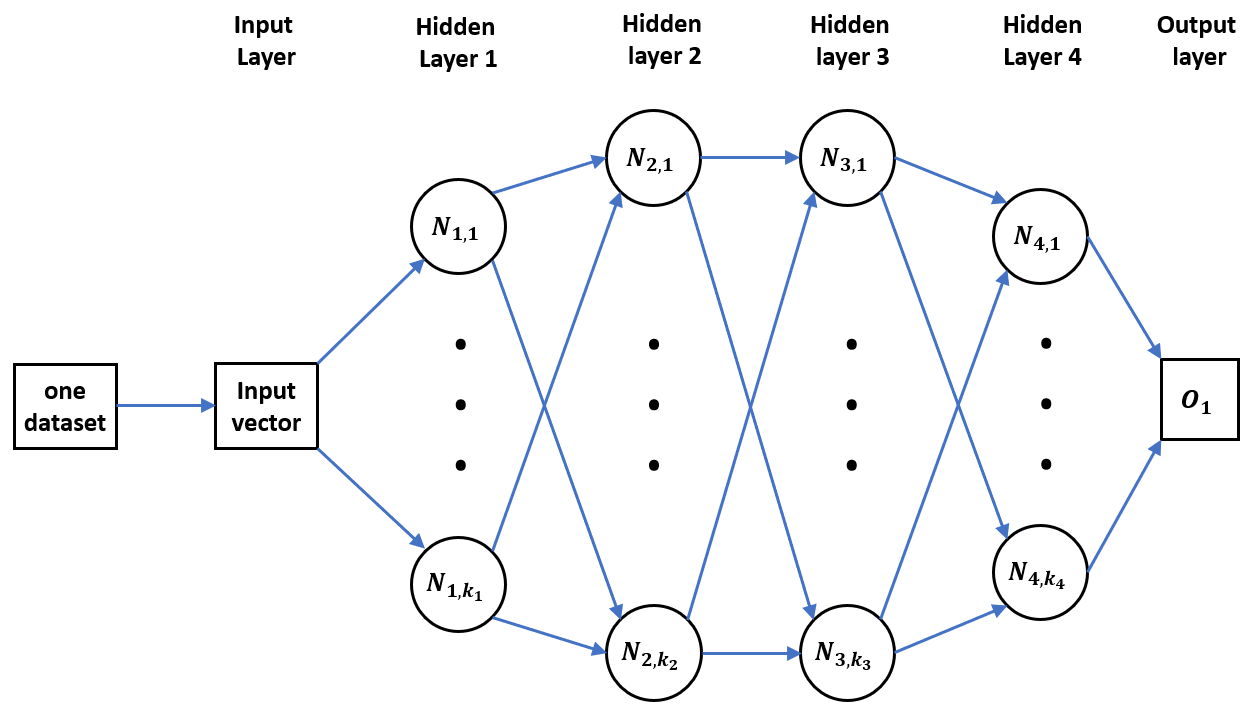}
	\caption{An illustration of a typical ST-DNN. Only one task (data set) is trained in this network.  Four hidden layers are included, $k_{i}$ (i = 1, 2, 3, 4) represents the number of neurons in the $i$th hidden layer and $N_{i,j}$ is the $j$th neuron in the $i$th hidden layer. Here, $O_{1}$ is the single output for the  task.}
	\label{fig:st-dnn}
\end{figure}

ST-DNN is a regular deep learning algorithm. It only takes care of one single prediction task. Therefore, it only learns from one specific training dataset. A typical four-layer ST-DNN is showed in figure ~\ref{fig:st-dnn}, where $N_i$ (i = 1, ..., 4), represents the number of neurons in the $i$th hidden layer.

\subsection{Multitask deep neural network (MT-DNN)}

\paragraph{}
The multitask (MT) learning technique has achieved much success in qualitative Merck and Tox21 prediction challenges \cite{capuzzi2016qsar,ramsundar2017multitask,wenzel2019predictive,ye2018integrated}. In the MT framework, multiple tasks share the same hidden layers.  However,  the output layer is attached to different tasks. This framework enables the neural network to learn all the data simultaneously for different tasks. Thus, the commonalities and differences among various datasets can be exploited. It has been showed that MT learning typically can improve the prediction accuracy of relatively small datasets if it combines with relatively larger datasets in its training.

\begin{figure}[h]
	\includegraphics[width=0.8\textwidth]{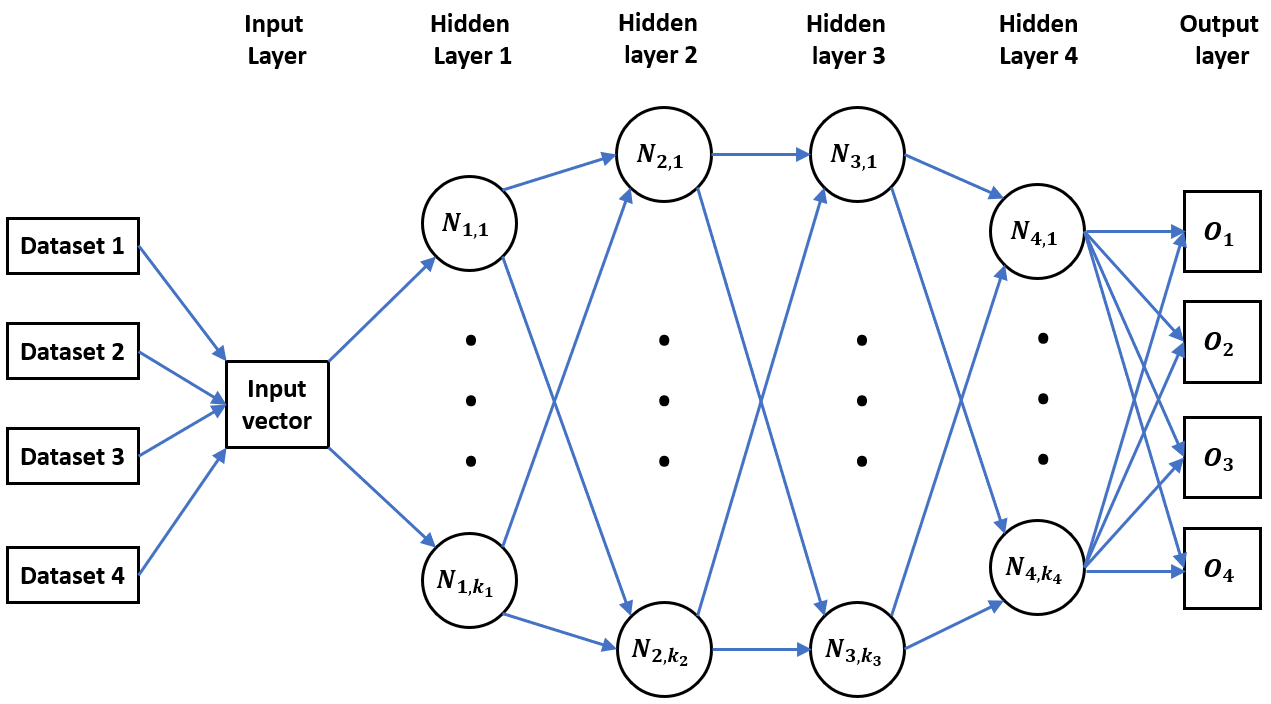}
	\caption{An illustration of a typical MT-DNN training four tasks (datasets) simultaneously. Four hidden layers are included in this network, $k_i$ (i = 1, 2, 3, 4) represents the number of neurons in the $i$th hidden layer and $N_{i,j}$ is the $j$th neuron in the $i$th hidden layer. Here $O_1$ to $O_4$ represent four predictor outputs for four tasks.}
	\label{fig:mt-dnn}
\end{figure}

\paragraph{}
Figure ~\ref{fig:mt-dnn} is an illustration of a typical four-layer MT-DNN for training four different tasks simultaneously. Suppose there are totally $T$ tasks and the training data for the $t$th task are $(X_i^t,y_i^t)_{i=1}^{N_t}$, where $t = 1,\dots, T$, $i = 1,\dots,N_t$, $N_t$ is the number of samples in the $t$th task, and $X_i^t$ is the feature vector for the $i$th sample in the $t$th task, $y_i^t$ is the label value of the $i$th sample in the $t$th task, respectively. The purpose of MT learning is to simultaneously minimize the loss function:
\begin{center}
${\rm argmin}\sum_{t=1}^T\sum_{i=1}^{N_t}{L}{(y_i^t,f^t (X_i^t,\theta^t))}$
\end{center}
where $f^t$ is the prediction for the $i$th sample in the $t$th task by our MT-DNN, which is a function of the feature vector $X_i^t$, $L$ is the loss function,   and $\theta^t$ is the collection of machine learning hyperparameters. A popular cost function for regression is the mean squared error, which can be defined as:
\begin{center}
${L}{(y_i^t,f^t (X_i^t,\theta^t))}=\frac{1}{N_t}\sum_{i=1}^{N_t}{(y_i^t-f^t (X_i^t,\theta^t))}{^2}$. 
\end{center}

\paragraph{} 
In this study, MT learning technology is applied to the toxicity prediction. The ultimate goal of this MT learning is to potentially improve the overall performance of multiple toxicity prediction models, especially for the smallest dataset that performs relatively poorly in the ST-DNN. More concretely, it is reasonable to assume that different toxicity indexes share a common pattern so that these different tasks can be trained simultaneously when their feature vectors are constructed in the same manner. For our toxicity prediction, four different tasks (LD$_{50}$, IGC$_{50}$, LC$_{50}$, LC$_{50}$-DM data sets)  are trained together. This leads to four output neurons in the output layer (See $O_1$ to $O_4$ in Figure ~\ref{fig:mt-dnn}), with each neuron being specific to one of  four tasks.

\subsection{Hyperparameters}

\paragraph{Ensemble hyperparameters.}
Both RF and GBDT were implemented by the scikit-learn package (version 0.20.1) \cite{pedregosa2011scikit}. In this work, there are a total of 23 datasets with their training data size varying from 94 to 8199. RF has been showed to be consistent and robust with various datasets. However, if its parameters are carefully tuned based on the size of a given training set, GBDT can attain better performance than RF does in most cases. For all experiments in this work, the most essential parameters of GBDT are chosen as learning rate = 0.01, min\_samples\_split = 3, max\_features=sqrt. Detail values of other parameters are  given in Table \ref{table:gbdt-para}.


\begin{table}[h]
	\centering
	\begin{tabular}{|c|P{6cm}|P{6cm}|}
		\rowcolor{Lightgreen}
		\hline

		\textbf{Training-set size} & \textbf{RF parameters} & \textbf{GBDT parameters} \\
		\hline

		$<$800 & \multirow{9}{6cm}{n\_estimators=1000, criterion=`mse', max\_depth=None, min\_samples\_split=2, min\_samples\_leaf=1, min\_weight\_fraction\_leaf=0.0} & \multicolumn{1}{m{6cm}|}{n\_estimators=2000, max\_depth=9, min\_samples\_split=3, learning\_rate=0.01, subsample=0.1, max\_features='sqrt'} \\
		\cline{1-1} \cline{3-3}

		800 to 5000 & & \multicolumn{1}{m{6cm}|}{n\_estimators=10000, max\_depth=7,min\_samples\_split=3, learning\_rate=0.01, subsample=0.3, max\_features='sqrt'} \\
		\cline{1-1} \cline{3-3}
		
		5000 to 10000 & & \multicolumn{1}{m{6cm}|}{n\_estimators=20000, max\_depth=7,min\_samples\_split=3, learning\_rate=0.01, subsample=0.3, max\_features='sqrt'} \\
		\hline

	\end{tabular}
	\caption{RF and GBDT parameters for different training-set sizes.}
	\label{table:gbdt-para}
\end{table}

\paragraph{Network hyperparameters.}
Since the numbers of features differ much in different 2D fingerprints, different network architectures have to be adopted. For example, Estate 1 fingerprint  has only 79 bits. Therefore  a 4-layer network with the number of neurons  in various hidden layers are chosen as   500, 1000, 1500, and 500. However,  the Daylight fingerprint has as many as 2048 features, and thus a much larger network is needed. The network for this fingerprint still has 4 layers but there are 3000, 2000, 1000, and 500 neurons in the first, second, third and fourth hidden layer, respectively. Other network parameters are as followed: the optimizer is  stochastic gradient descent (SGD) with momentum of 0.5. 2000 epochs were run for all the networks.  Mini-batch size is set to 4.  The learning rate is set to 0.01 in the first 1000 epochs and  0.001 for the rest epochs. Our tests indicate that adding a dropout or using $L_2$ decay does not necessarily improve the accuracy, and thus, we omit these two techniques. All the network hyperparameters are summarized in Table ~\ref{table:net-para}. These hyperparameters are applied to both ST-DNN and MT-DNN. All the DNN training is performed by Pytorch (version 1.0) \cite{paszke2017pytorch}.

\begin{table}[h]
	\centering
	\begin{tabular}{|P{2cm}|P{1.5cm}|P{2cm}|P{3cm}|P{2cm}|P{1cm}|P{2cm}|}
		\hline
		\rowcolor{Lightgreen}
		\textbf{Fingerprint} & \textbf{Number of features} & \textbf{Number of hidden layers} & \textbf{Number of neurons in each hidden layer} & \textbf{Optimizer} & \textbf{Mini-batch} & \textbf{Learning rate}\\
		\hline
		Estate1 & 79 & \multirow{3}{*}{4} & \multirow{2}{3cm}{500,1000,1500,500} & \multirow{1}{2cm}{SGD with a momentum of 0.5} & \multirow{3}{*}{4} & \multirow{3}{2cm}{First 1000: 0.01; Then: 0.001}\\
		\cline{1-2}
		Estate2 & 79 & & & & &\\
		\cline{1-2} \cline{4-4}
		Daylight & 2048 & & 3000,2000,1000,500 & & & \\
		\hline
	\end{tabular}
	\caption{The network hyperparameters for both ST-DNN and MT-DNN.}
	\label{table:net-para}
\end{table}

\section{Results}

\subsection{Toxicity prediction}

\paragraph{}
Four toxicity datasets were studied in our work, namely oral rat LD$_{50}$ (LD$_{50}$), 40 h Tetrahymena pyriformis IGC$_{50}$ (IGC$_{50}$), 96 h fathead minnow LC$_{50}$ (LC$_{50}$), and 48 h Daphnia magna LC$_{50}$ (LC$_{50}$-DM). Among them, LD$_{50}$ measures the amount of chemicals that can kill half of rats when orally ingested. IGC$_{50}$ records the 50\% growth inhibitory concentration of Tetrahymena pyriformis organism after 40 h. LC$_{50}$ reports at the concentration of test chemicals in water in milligrams per liter that cause 50\% of fathead minnows to die after 96 h. The last one is LC$_{50}$-DM, which represents the concentration of test chemicals in water in milligrams per liter that cause 50\% Daphnia maga to die after 48 h. The unit of toxicity reported in these four datasets is -log$_{10}$ mol/L. All of them are accessible from the recent publications \cite{martin2016user,akers1999structure,zhu2008combinatorial} and the public database (https://www.epa.gov/chemical-research/toxicity-estimation-software-tool-test). The sizes of these four datasets vary from 353 to 7413 (See Table ~\ref{table:tox-dataset}), which raises a challenge for a predictive model to achieve  a consistent accuracy and robustness.

\begin{table}[h]
	\centering
	\begin{tabular}{|c|c|c|c|c|c|}

		\hline
		\rowcolor{Lightgreen}
		\textbf{Data set} & \textbf{Total size} & \textbf{Train set size} & \textbf{Test set size} & \textbf{Max value} & \textbf{Min value} \\
		\hline
		LD$_{50}$ & 7413 & 5931 & 1482 & 7.201 & 0.291 \\
		\hline
		IGC$_{50}$ & 1792 & 1434 & 358 & 6.36 & 0.334 \\
		\hline
		LC$_{50}$ & 823 & 659 & 164 & 9.261	& 0.037 \\
		\hline
		LC$_{50}$-DM & 353 & 283 & 70 & 10.064 & 0.117\\
		\hline
	\end{tabular}
	\caption{The quantitative summary of  four toxicity datasets. The original datasets and prediction results are available at  
	\href{https://www.epa.gov/chemical-research/toxicity-estimation-software-tool-test}{https://www.epa.gov/chemical-research/toxicity-estimation-software-tool-test}.
	}
	\label{table:tox-dataset}
\end{table}

\subsubsection{The performance of ensemble methods}

\paragraph{}
Because it is easy to implement and fast to train, two ensemble methods, RF and GBDT, were first tested. Since four datasets have very different sizes, different numbers of estimators in RF and GBDT models should be used. Specifically, for two relatively small sets, LC$_{50}$ and LC$_{50}$-DM, the numbers of estimators are set to 2000. For IGC50, 10000 estimators are used. For the largest set LD$_{50}$, we have used 20000 estimators.

\begin{figure}[h]
	\includegraphics[width=\textwidth]{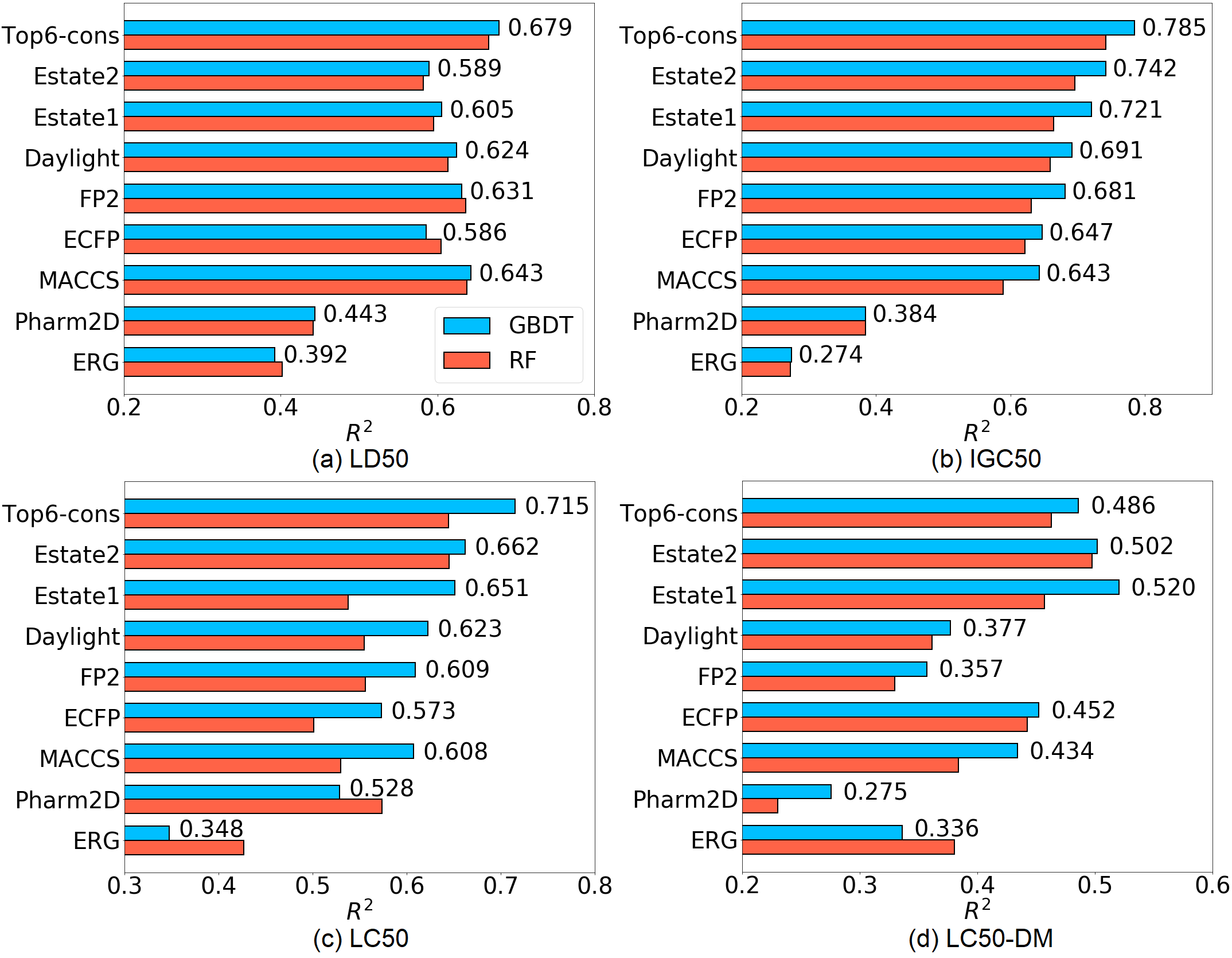}
	\caption{The $R^2$ on LD$_{50}$, IGC$_{50}$, LC$_{50}$, LC$_{50}$-DM test sets yielded by eight fingerprints and the consensuses of the top 6 features.  Two ensemble methods were adopted (GBDT: blue, RF: red).  The values shown in the figure are the $R^2$ of GBDT.}
	\label{fig:tox-gbdt}
\end{figure}

\paragraph{}
The accuracy is in term of the square of Pearson correlation coefficient ($R^2$). Overall, GBDT's performance is always better than that of RF, which agrees with early publication \cite{wu2018quantitative}. Among all  the eight fingerprints we tested, Estate2, Estate1, Daylight, FP2, ECFP and MACCS usually work well on these four sets. Thus the consensus of these six fingerprints was also considered (``Top 6-cons'' in Figure ~\ref{fig:tox-gbdt}). The consensus model typically gives rise to a further improvement over all single fingerprints in most cases.

\paragraph{(a)LD$_{50}$ test set.}
LD$_{50}$  dataset is the largest set having as many as 7413 compounds. However, a higher experimental uncertainty of the values in this set makes this set relatively difficult to predict (See "Max value" and "Min value" in Table ~\ref{table:tox-dataset}). In our GBDT model, the best single fingerprint (MACCS) yields an $R^2$ of 0.643, while the consensus of the top 6 fingerprints increases $R^2$ to 0.679.

\paragraph{(b) IGC$_{50}$ test set.}
IGC$_{50}$ set is the second largest set (1792 compounds) among the four sets we investigated. As indicated in Table ~\ref{table:tox-dataset}, the diversity of molecules of in this set is the lowest among the four sets, which means the prediction should be some easier. Our results show that Estate2 is the best single fingerprint with an $R^2$ of 0.742, and the consensus of the top 6 fingerprints leads to an $R^2$ of 0.785.

\paragraph{(c) LC$_{50}$ test set.}
LC$_{50}$ set is a relative smaller set (823 compounds). By employing GBDT model, estate2 fingerprint achieves the top performance, which yields an $R^2$ of 0.662. The consensus of the top 6 fingerprint improves the $R^2$ to 0.715.


\paragraph{(d) LC$_{50}$-DM test set.}
Among the four sets, LC$_{50}$-DM test set is the smallest one with only 283 training molecules and 70 test molecules, which is troublesome to build a robust model. Therefore, our model just generates moderate results. Specifically, the best single fingerprint Estate1 only has an $R^2$ of 0.520. The consensus model even ruins the $R^2$ a little bit with an $R^2$ as low as 0.486. Similar difficulty is also faced by other recent work, such as the $R^2$ of the 3D-topology based GBDT model only reaches 0.505 \cite{wu2018quantitative}. Thus, there is a need for multitask deep learning when dealing with such a small dataset.

\subsubsection{The performance of single-task and multitask deep learning}

\paragraph{}
On average, Estate2, Estate1, and Daylight are the top three fingerprints when using GBDT models in all the four sets. Thus, these three fingerprints were picked up to perform higher-level ST-DNN and MT-DNN.

\paragraph{}
Since the lengths of the three fingerprints differ much, different DNN architectures are needed. 
Four hidden layers with 500, 1000, 1500, and 500 neurons are used for Estate1 and Estate2, whose  fingerprints have 79 features.  Four hidden layers with 3000, 2000, 1000, and 500 neurons are used for Daylight, whose  fingerprint has 2048 bits.

\paragraph{}
The pattern of  ST-DNN results is similar to that of  GBDT results. On  four data sets, a ST-DNN consensus model yields an average $R^2$ of 0.658 (0.632, 0.791, 0.687, and 0.523 respectively).  As a comparison, the average $R^2$ by a GBDT consensus model is 0.666 (0.679, 0.785, 0.715, and 0.486 respectively). However, the performance can be largely enhanced by the multitask strategy because the two relatively smaller sets LC$_{50}$ and LC$_{50}$-DM can benefit much from  two larger sets LD$_{50}$ and IGC$_{50}$. As shown in Table ~\ref{table:tox-dnn}, while the MT-DNN model seldom changes the performance on LD$_{50}$ and IGC$_{50}$, it gives rise to a dramatic improvement on LC$_{50}$ and LC$_{50}$-DM, especially on LC$_{50}$-DM.  The consensus lifts the $R^2$ result from 0.523 to 0.725. 

\begin{table}[h!]
	\centering
	\begin{tabular}{|ccccc|}
		\hline
		\rowcolor{Lightgreen}
		\textbf{Method} & \textbf{$R^2$ of LD$_{50}$} & \textbf{$R^2$ of IGC$_{50}$} & \textbf{$R^2$ of LC$_{50}$} & \textbf{$R^2$ of LC$_{50}$-DM} \\
		\hline
		Estate2 ST-DNN	& 0.484	& 0.715	& 0.569	& 0.433 \\
		Estate2 MT-DNN	& 0.489	& 0.696	& 0.660	& 0.623 \\
		Estate1 ST-DNN	& 0.569	& 0.733	& 0.650	& 0.601 \\
		Estate1 MT-DNN	& 0.566	& 0.735	& 0.694	& 0.684 \\
		Daylight ST-DNN & 0.619 & 0.701 & 0.570 & 0.346 \\
		Daylight MT-DNN & 0.617 & 0.717 & 0.724 & 0.694 \\
		Consensus ST-DNN & 0.632 & 0.791 & 0.687 & 0.523 \\
		Consensus MT-DNN & 0.639 & 0.794 & 0.765 & 0.725 \\
		\hline
	\end{tabular}
	\caption{The $R^2$ of ST-DNN and MT-DNN based on the top 3 fingerprints in GBDT (Estate2, Estate1, Daylight) and their consensuses.}
	\label{table:tox-dnn}
\end{table}

\begin{table}[h]
	\centering
	\begin{tabular}{|cccc|}
		
		\hline 
		
		\rowcolor{Lightgreen}
		\multicolumn{4}{|c|}{\textbf{LD$_{50}$}} \\
		\hline
		Method & $R^2$ & RMSE & Coverage \\
		\hline
		\textbf{The present 2D MT-DNN consensus} & \textbf{0.639} & \textbf{0.549} & \textbf{1.000} \\
		\textbf{The present 2D GBDT consensus} & \textbf{0.679} & \textbf{0.580} & \textbf{1.000} \\
		\hline
		Hierarchical \cite{martin2016user} & 0.578 & 0.650 & 0.876 \\
		FDA	\cite{martin2016user} & 0.557	& 0.657	& 0.984 \\
		Nearest neighbor \cite{martin2016user} & 0.557 & 0.656 & 0.993 \\
		T.E.S.T consensus \cite{martin2016user} & 0.626 & 0.594 & 0.984 \\
		\hline
		3D MT-DNN consensus \cite{wu2018quantitative} & 0.653 & 0.568 & 0.997 \\
		\hline
		\rowcolor{Lightgreen}
		\multicolumn{4}{|c|}{\textbf{IGC$_{50}$}} \\
		\hline
		Method & $R^2$ & RMSE & Coverage \\
		\hline
		\textbf{The present 2D MT-DNN consensus} & \textbf{0.794} & \textbf{0.457} & \textbf{1.000} \\
		\textbf{The present 2D GBDT consensus} & \textbf{0.785} & \textbf{0.457} & \textbf{1.000} \\
		\hline
		Hierarchical \cite{martin2016user} & 0.719 & 0.539 & 0.933 \\
		FDA	\cite{martin2016user} & 0.747	& 0.489	& 0.978 \\
		Group contribution \cite{martin2016user} & 0.682 & 0.575 & 0.955 \\
		Nearest neighbor \cite{martin2016user} & 0.600 & 0.638 & 0.986 \\
		T.E.S.T consensus \cite{martin2016user} & 0.764 & 0.475 & 0.983 \\
		\hline
		3D MT-DNN consensus\cite{wu2018quantitative} & 0.802 & 0.438 & 1.000 \\
		\hline
		\rowcolor{Lightgreen}
		\multicolumn{4}{|c|}{\textbf{LC$_{50}$}} \\
		\hline
		Method & $R^2$ & RMSE & Coverage \\
		\hline
		\textbf{The present 2D MT-DNN consensus} & \textbf{0.765} & \textbf{0.718} & \textbf{1.000} \\
		\textbf{The present 2D GBDT consensus} & \textbf{0.715} & \textbf{0.783} & \textbf{1.000} \\
		\hline
		Hierarchical \cite{martin2016user} & 0.710 & 0.801 & 0.951 \\
		Single model \cite{martin2016user} & 0.704 & 0.803 & 0.945 \\
		FDA	\cite{martin2016user} & 0.626 & 0.915 & 0.945 \\
		Group contribution \cite{martin2016user} & 0.686 & 0.810 & 0.872 \\
		Nearest neighbor \cite{martin2016user} & 0.667 & 0.876 & 0.939 \\
		T.E.S.T consensus \cite{martin2016user} & 0.728 & 0.768 & 0.951 \\
		\hline
		3D MT-DNN consensus\cite{wu2018quantitative}	& 0.789 & 0.677 & 1.000 \\
		\hline
		\rowcolor{Lightgreen}
		\multicolumn{4}{|c|}{\textbf{LC$_{50}$-DM}} \\
		\hline
		Method & $R^2$ & RMSE & Coverage \\
		\hline
		\textbf{The present 2D MT-DNN consensus} & \textbf{0.725} & \textbf{0.935} & \textbf{1.000} \\
		\textbf{The present 2D GBDT consensus} & \textbf{0.486} & \textbf{1.239} & \textbf{1.000} \\
		\hline
		Hierarchical \cite{martin2016user} & 0.695 & 0.979 & 0.886 \\
		Single model \cite{martin2016user} & 0.697 & 0.993 & 0.871 \\
		FDA	\cite{martin2016user} & 0.565	& 1.190	& 0.900 \\
		Group contribution \cite{martin2016user} & 0.671 & 0.803 & 0.657 \\
		Nearest neighbor \cite{martin2016user} & 0.733 & 0.975 & 0.871 \\
		T.E.S.T consensus \cite{martin2016user} &	0.739 & 0.911 & 0.900 \\
		\hline
		3D MT-DNN consensus\cite{wu2018quantitative}	& 0.678	& 0.978	& 1.000\\
		\hline
	\end{tabular}
	\caption{Comparison to other toxicity prediction methods. The prediction results for Hierarchical, Single model, FDA, Group contribution, Nearest neighbor, and T.E.S.T consensus are available in reference 44 and at \href{https://www.epa.gov/chemical-research/toxicity-estimation-software-tool-test}{https://www.epa.gov/chemical-research/toxicity-estimation-software-tool-test}.
	}
	\label{table:tox-compare}
\end{table}

\subsubsection{Systematic comparison with other toxicity predictions}
\paragraph{}
A systematic comparison with other  methods was provided in Table ~\ref{table:tox-compare}. The same datasets are also used to develop the Toxicity Estimation Software Tool (T.E.S.T). So many related results can be found in its user's guide \cite{martin2016user}, including hierarchical, single model, FDA, group contribution, nearest neighbor, and T.E.S.T consensus.

\paragraph{}
Since T.E.S.T is also based on  2D descriptors, the comparison between the results from  the present models and T.E.S.T can largely reflect the predictive power of the present models. As shown in Table ~\ref{table:tox-compare}, on the LD$_{50}$, IGC$_{50}$, LC$_{50}$ sets, the present MT-DNN consensus always leads to a higher $R^2$ than T.E.S.T consensus. Especially, on the IGC$_{50}$ and LC$_{50}$ sets, the present MT-DNN consensus models largely beat T.E.S.T (0.794 vs 0.764 and 0.765 vs 0.728), and the present GBDT results quite outperform T.E.S.T (0.679 vs 0.626) on the LD$_{50}$ set. Even on the LC50-DM set, because the training set is so small (283), ensemble methods (RF and GBDT) and DNN methods are not suitable for it: $R^2$ of ST-DNN and GBDT are, respectively, 0.486 and 0.523. However, the $R^2$  of MT-DNN is as high as 0.725 for LC$_{50}$-DM dataset,  which is quite comparable to the T.E.S.T  result with an $R^2$ of 0.739.


\paragraph{}
2D MT-DNN consensus has an average $R^2$ of 0.731 for these four datasets, while the average of T.E.S.T model is 0.714, and the recent 3D structure-based topological MT-DNN consensus result is also 0.731 \cite{wu2018quantitative}. These results confirm that 2D fingerprints integrated with MT-DNN model surpass the previous 2D models and are as good as  the recent 3D structure-based topological model \cite{wu2018quantitative}.

\subsection{Aqueous solubility (Log S) }

\paragraph{}
For Log S, following the previous literature \cite{wu2018topp,hou2004adme}, we test Klopman's test set \cite{klopman1992estimation} with the original train set. The unit of Log P in these sets is log unit. Since the size of the training set is 1290, 10000 estimators were used in the GBDT model.

\begin{table}[h!]
	\centering
	\begin{tabular}{|c|c|}
		\hline
		\rowcolor{Lightgreen}
		\textbf{Training set} & \textbf{Klopman's test set} \\
		\hline
		1290 & 21 \\
		\hline
	\end{tabular}
	\caption{The sizes of Log S training set and Klopman's test set.}
	\label{table:logs-dataset}
\end{table}

\begin{table}[h!]
	\centering
	\begin{tabular}{|ccc|}
		\hline
		\rowcolor{Lightgreen}
		\textbf{Fingerprint} & \textbf{$R$} & \textbf{RMSE} \\
		\hline
		\textbf{Cons-top 3}	& \textbf{0.955} & \textbf{0.648} \\
		\textbf{Cons-top 6} & \textbf{0.944} & \textbf{0.684} \\
		MACCS &	0.958 & 0.664 \\
		Estate1	& 0.932	& 0.791 \\
		Daylight & 0.923 & 0.780 \\
		FP2	& 0.908	& 0.853 \\
		ECFP & 0.904 & 0.875 \\
		Estate2	& 0.897	& 0.907 \\
		Pharm2D	& 0.832	& 1.114 \\
		ERG	& 0.811	& 1.202 \\
		\hline
	\end{tabular}
	\caption{The $R$ and RMSE of predicting Log S by   eight fingerprints and the consensuses of the top 3 and top 6 on Klopman's test set.}
	\label{table:logs-results}
\end{table}

\begin{table}[h]
	\centering
	\begin{tabular}{|ccc|}
		\hline
		\rowcolor{Lightgreen}
		\textbf{Method} & \textbf{$R$} & \textbf{RMSE} \\
		\hline
		\textbf{Cons-top 3}	& \textbf{0.955} & \textbf{0.648} \\
		\textbf{Cons-top 6} & \textbf{0.944} & \textbf{0.684} \\
		\hline
		MT-ESTD$^+$-1 (3D) \cite{wu2018topp} & 0.94	& 0.69 \\
		Drug-LOGS (2D) \cite{hou2004adme} & 0.94 & 0.64 \\
		Klopman MLR (2D) \cite{klopman1992estimation} & 0.92	& 0.86 \\
		\hline
	\end{tabular}
	\caption{Comparison of prediction results on the Log S data set.}
	\label{table:logs-compare}
\end{table}

\paragraph{}
In the Log S test, the top 6 fingerprints are MACCS, FP2, Daylight, Estate1, Estate2, and ECFP, which perform much better than the other two fingerprints, Pharm2D and ERG. The consensuses of the top 6 fingerprints results in $R$ and RMSE of 0.944 and 0.684, respectively. The consensus of top 3 is even better, which improves $R$ and RMSE to 0.955 and 0.648 (See Table ~\ref{table:logs-results}). A systematic comparisons to other methods are included in Table ~\ref{table:logs-compare}. It indicates the present method outperforms all other state-of-the-art 3D and 2D methods.

\subsection{Partition coefficient (Log P)}

\paragraph{}
Three Log P data sets were tested using the GBDT model. The training set has 8199 molecules, which was originally compiled by Cheng {\it et al.} \cite{cheng2007computation}. There are three test sets, namely FDA \cite{cheng2007computation}, Star \cite{avdeef2012absorption}, and Non-star \cite{avdeef2012absorption} respectively, which are given in Table ~\ref{table:logp-dataset}. The Log P in these sets is by the unit of log$_{10}$ mol/L. Due to the size of the training set, 20000 estimators are used in the GBDT model.

\begin{table}[h]
	\centering
	\begin{tabular}{|c|c|c|c|}
		\hline
		\rowcolor{Lightgreen}
		 & \multicolumn{3}{|c|}{\textbf{Test set}} \\
		\cline{2-4}
		\cellcolor{Lightgreen}\multirow{-2}{*}{\textbf{Training set}}& \textbf{FDA} & \textbf{Star} & \textbf{Non-star} \\
		\hline
		8199 & 406 & 223 & 43 \\
		\hline
	\end{tabular}
	\caption{The sizes of Log P training set and test sets.}
	\label{table:logp-dataset}
\end{table}
\begin{figure}[ht!]
	\includegraphics[width=\textwidth]{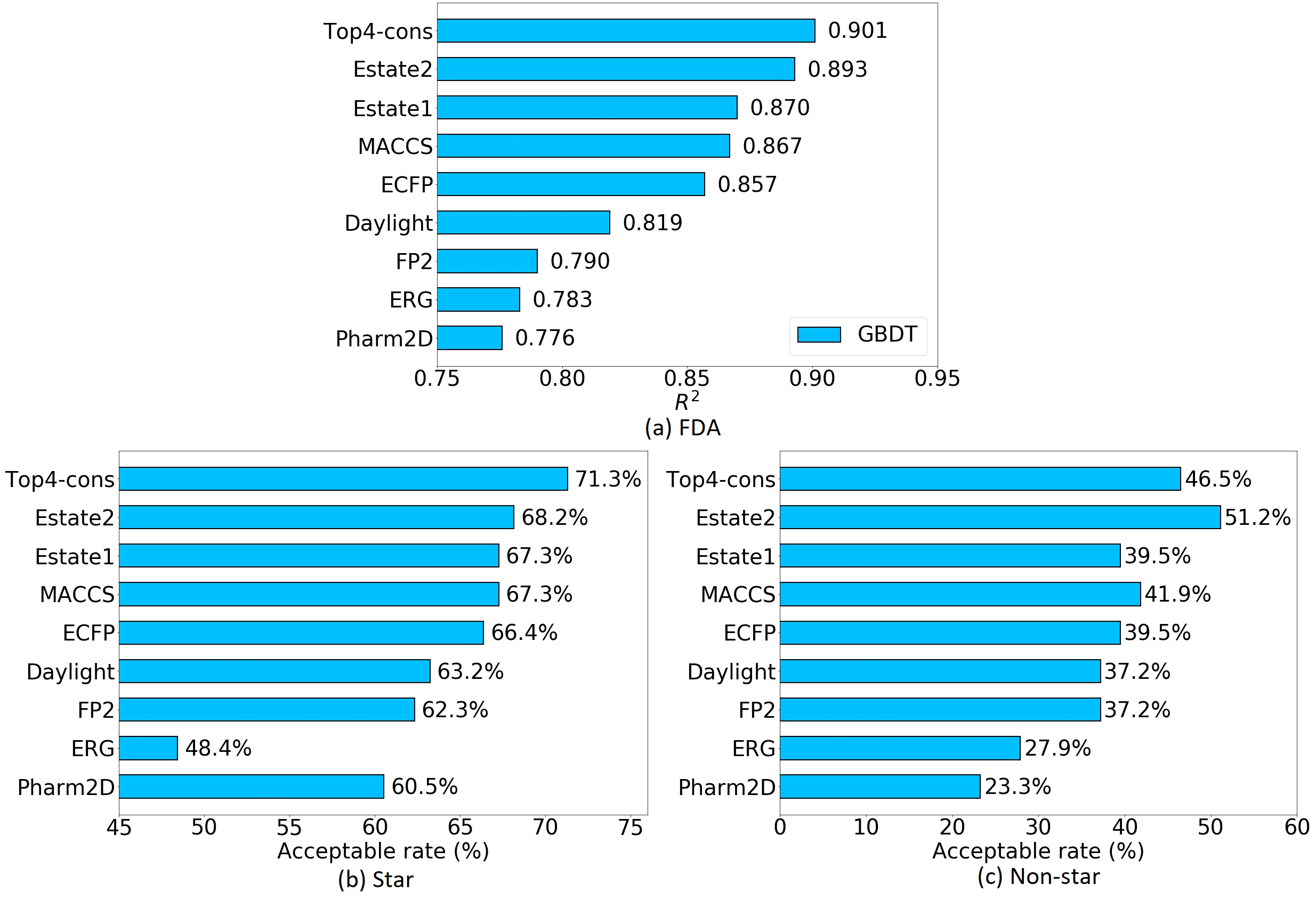}
	\caption{ The performance of eight fingerprints and the consensuses of the top 4 on the FDA, Star and Non-star data sets of Log P. To be consistent with previous results, on the FDA set, $R^2$ is given, while on star and non-star datasets,  acceptable rate is given.}
	\label{fig:logp-result}
\end{figure}

\paragraph{}
In order to easily compare to the earlier literatures, accuracy on these three test sets are reported by $R^2$ or acceptable rate. The acceptable rate here is defined as the percentage of molecules within error range < 0.5 \cite{mannhold2009calculation}. Of all the three sets, the 2D fingerprints of Estate2, Estate1, MACCS, and ECFP are always the top 4.  
 The consensuses of the top 4 fingerprints produce $R^2$ up to 0.901 on the FDA set and attain an acceptable rate on Star set at 71.3\%. On the Non-star set, the top 4 consensus is somehow worse than the best single fingerprint Estate1 but it is still in the second place with an acceptable rate of 46.5\% (See Figure ~\ref{fig:logp-result}).

\paragraph{}
A detailed comparison with other Log P prediction methods was shown in Table \ref{table:logp-fda-compare}. On the FDA data set, GBDT-ESTD$^+$-2-AD \cite{wu2018topp} and MT-ESTD-1 \cite{wu2018topp} are based on 3D descriptors. GBDT-ESTD$^+$-2-AD model includes some molecules from the NIH-dataset in its training set. Therefore, its performance is slightly better than the present one. 
The 2D method ALOGPS \cite{cheng2007computation} also performs slightly better (0.908 vs 0.901) than the present one. However, a previous study \cite{hou2004adme} has pointed out that for the PHYSPROP database \cite{howard1999physical},  the training set of ALOGPS actually contains all of the compounds in the FDA set.  It is unclear how well it will perform if the overlapping compounds are removed from the training set. Unlike ALOGPS, XLOGP3's training data is completely independent of the test set \cite{cheng2007computation}. In this case, the present prediction is more accurate  than that of XLOGP3  (0.901 vs 0.872).

\begin{table}[h!]
	\centering
	\begin{tabular}{|ccc|}
		\hline
		\rowcolor{Lightgreen}
		\textbf{Method} & \textbf{$R^2$} & \textbf{RMSE} \\
		\hline
		GBDT-ESTD$^+$-2-AD (2D+3D) \cite{wu2018topp} & 0.935 & 0.51 \\
		MT-ESTD-1 (3D) \cite{wu2018topp} & 0.920 & 0.57 \\
		ALOGPS (2D but the training set contains test set) \cite{cheng2007computation} & 0.908	& 0.60 \\
		\textbf{Our Cons-top 4 (2D)} & \textbf{0.901} & \textbf{0.63} \\
		XLOGP3 (2D) \cite{cheng2007computation} & 0.872	& 0.72 \\
		XLOGP3-AA (2D) \cite{cheng2007computation} & 0.847 & 0.80 \\
		CLOGP (2D) \cite{cheng2007computation} & 0.838	& 0.88 \\
		TOPKAT (2D) \cite{cheng2007computation}	& 0.815	& 0.88 \\
		ALOGP98 (2D) \cite{cheng2007computation} & 0.80	& 0.90 \\
		KowWIN (2D) \cite{cheng2007computation} & 0.771	& 1.10 \\
		HINT (2D) \cite{cheng2007computation} & 0.491 & 1.93 \\
		\hline
	\end{tabular}
	\caption{Comparison of  Log P predictions  on the FDA set. }
	\label{table:logp-fda-compare}
\end{table}

\begin{table}[h]
	\centering
	\begin{tabular}{|ccccccc|}
		\hline
		\rowcolor{Lightgreen}
		& \multicolumn{3}{|c|}{\textbf{Star set (N=223)}} & \multicolumn{3}{|c|}{\textbf{Non-star set (N=43)}} \\
		\hline
		& \multicolumn{2}{c}{\multirow{2}{3cm}{\% of Molecules within error range}} & & \multicolumn{2}{c}{\multirow{2}{3cm}{\% of Molecules within error range}} & \\
		& \multicolumn{2}{c}{} & & \multicolumn{2}{c}{} & \\
		\cline{2-3} \cline{5-6}
		Method	& $<$0.5 & $<$1 & RMSE & $<$0.5 & $<$1 & RMSE \\
		\hline
		AB/LogP	\cite{mannhold2009calculation} & 84 & 12 & 0.41 & 42 &	23 & 1.00 \\
		MT-ESTD$^+$-1-AD \cite{wu2018topp} & 77 & 16 & 0.49 & 49 & 19 & 0.98 \\
		S+logP \cite{mannhold2009calculation} & 76 & 22 & 0.45 & 40 & 35 & 0.87 \\
		ACD/logP \cite{mannhold2009calculation} & 75 & 17 & 0.50 & 44 & 32	& 1.00 \\
		CLOGP \cite{mannhold2009calculation} & 74 & 20	& 0.52 & 47	& 28 & 0.91 \\
		MT-ESTD-1 \cite{wu2018topp} & 72 & 18 & 0.55 & 33 & 28	& 1.01 \\
		ALOGPS \cite{mannhold2009calculation} & 71	& 23 & 0.53	& 42 & 30 & 0.82 \\
		\textbf{Our cons-top 4} & \textbf{71} & \textbf{18} & \textbf{0.625} & \textbf{47} & \textbf{16} & \textbf{1.233} \\
		MiLogP \cite{mannhold2009calculation} & 69 & 22 & 0.57 & 49 & 30 & 0.86 \\
		KowWIN \cite{mannhold2009calculation} & 68 & 21 & 0.64 & 40 & 30 & 1.05 \\
		TLOGP \cite{mannhold2009calculation} & 67 & 16	& 0.74	& 30 & 37 & 1.12 \\
		CSLogP \cite{mannhold2009calculation} & 66	& 22 & 0.65	& 58 & 19 & 0.93 \\
		SLIPPER-2002 \cite{mannhold2009calculation} & 62 & 22 & 0.80 & 35 & 23	& 1.23 \\
		XLOGP3 \cite{mannhold2009calculation} & 60	& 30 & 0.62	& 47 & 23 & 0.89 \\
		XLOGP2 \cite{mannhold2009calculation} & 57	& 22 & 0.87	& 35 & 23 & 1.16 \\
		QLOGP \cite{mannhold2009calculation} & 48 & 26	& 0.96 & 21	& 26 & 1.42 \\
		VEGA \cite{mannhold2009calculation} & 47 & 27 & 1.04 & 28 & 30	& 1.24 \\
		SPARC \cite{mannhold2009calculation} & 45 & 22	& 1.36 & 28	& 21 & 1.70 \\
		LSER \cite{mannhold2009calculation} & 44 & 26 & 1.07 & 35 & 16	& 1.26 \\
		CLIP \cite{mannhold2009calculation} & 41 & 25 & 1.05 & 33 & 9	& 1.54 \\
		MLOGP(Sim+)	\cite{mannhold2009calculation} & 38 & 30 & 1.26 & 26 & 28 & 1.56 \\
		HINTLOGP \cite{mannhold2009calculation} & 34 & 22 & 1.80 & 30 & 5 & 2.72 \\
		NC+NHET	\cite{mannhold2009calculation} & 29 & 26 & 1.35 & 19 & 16 & 1.71 \\
		\hline
	\end{tabular}
	\caption{Comparison of  Log P predictions of the Star and Nonstar sets.}
	\label{table:logp-star-nonstar-compare}
\end{table}

\paragraph{}
The present results on the Star and Non-star sets are also systematically compared with other stat-of-the-art models as shown in Table ~\ref{table:logp-star-nonstar-compare}. For the Star set, we achieve 71\% of total number of molecules having the predicted error less than 0.5 (acceptable rate 71\%). This result is quite satisfactory and is comparable to the 3D structure-based model developed by Wu {\it et al.} \cite{wu2018topp} with an acceptable rate of 72\% on the same training set (``MT-ESTD-1" in Table ~\ref{table:logp-star-nonstar-compare}). There are many commercial software packages developed to predict Log P such as AB/Log P \cite{mannhold2009calculation}, S/Log P \cite{mannhold2009calculation}, ACD/log P \cite{mannhold2009calculation}, etc. However, we cannot validate whether the training sets used in these software packages overlap with the Star set. It is more meaningful when comparing the present model to XLogP3 software  \cite{mannhold2009calculation} since its training dataset does not contain any molecules in the test set. Again, the present model outperforms XLogP3 package on the Star set with the acceptable rates being 71\% and 60\%, respectively. In the Non-star set, all of the published methods perform as accurate as those in the FDA and Star data set, since the structures in the Non-star set are relatively new and complex. Thus, our model also only achieves an acceptable rate of 47\%. However, it is still tied for the third place among all predictors. This result is even better than some 3D structure-based models, though RMSE is relatively high due to a few large outliers.

\subsection{Protein-ligand binding affinity prediction}

\subsubsection{The S1322 dataset}

\begin{figure}[ht!]
	\includegraphics[width=0.9\textwidth]{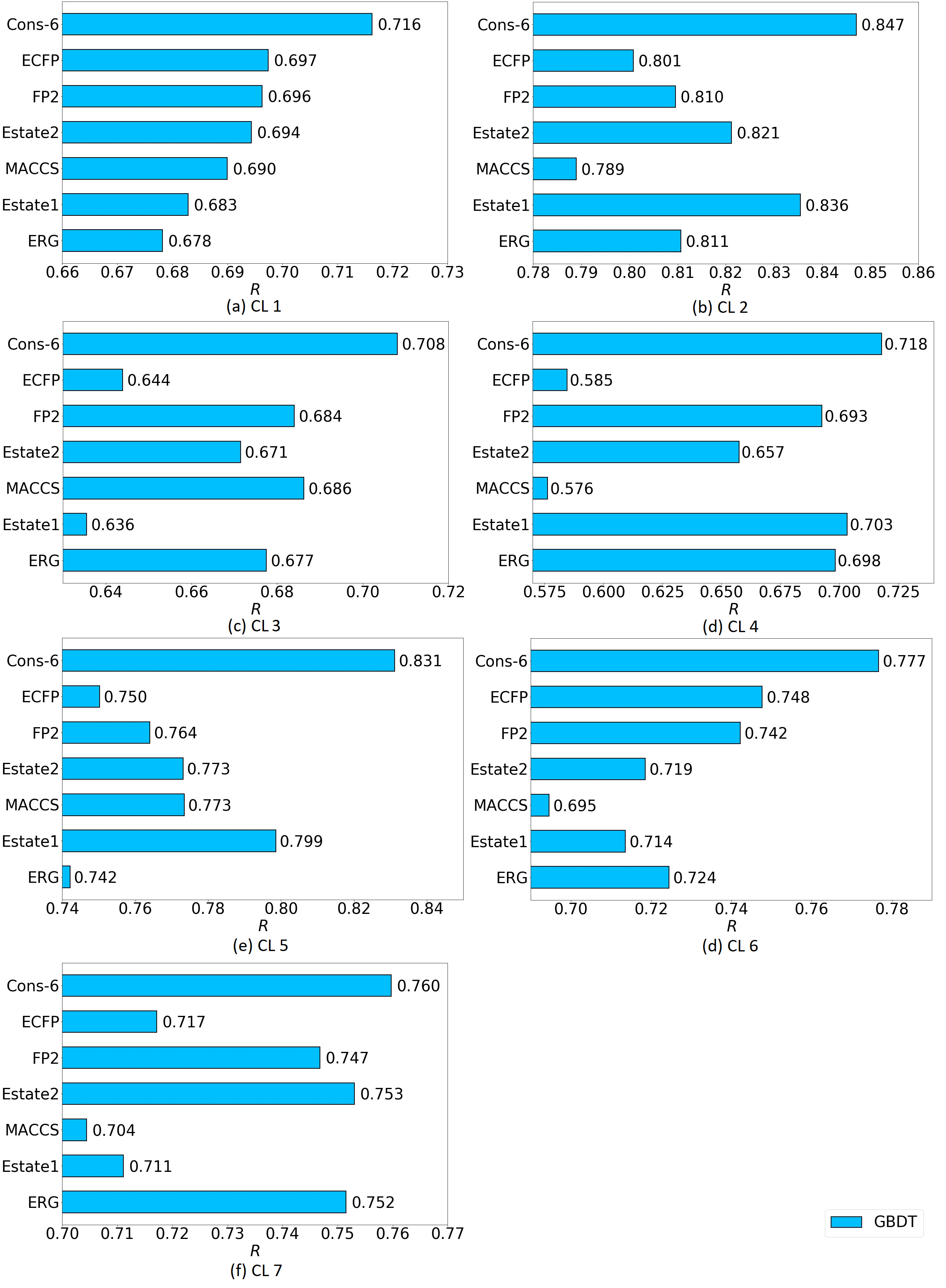}
	\caption{Pearson correlation coefficient ($R$) on the seven clusters of the S1322 data set yielded by the six fingerprints (ECFP, FP2, Estate 2, MACCS, Estate 1, ERG) and their consensuses.}
	\label{fig:s1322-bd}
\end{figure}

\paragraph{}
To assess the predictive power of  2D-fingerprint based models, two protein-ligand binding affinity datasets were investigated. The first one is denoted as the S1322 set. It is a high quality data set with 1322 protein-ligand complexes involving 7 protein clusters (labeled as  CL1, CL2,  $\cdots$, CL7). It is a subset of the refined set of PDBbind v2015 \cite{liu2014pdb}. The other dataset is PDBbind v2016 \cite{su2018comparative}, 
in which the refined set excluding the core set in PDBbind v2016 is used as a training data. The core set is a test set.
These two sets are summarized in Table ~\ref{table:bd-set}.

\begin{table}[ht!]
	\centering
	\begin{tabular}{|ccccccc|ccc|}
		\hline
		\rowcolor{Lightgreen}
		\multicolumn{7}{|c}{\textbf{S1322 set}} & \multicolumn{3}{|c|}{\textbf{PDBBind v2016 refined set}} \\
		\hline
		CL1 & CL2 & CL3 & CL4 & CL5 & CL6 & CL7 & refined set & training set & core set (test set) \\
		\hline
		333	& 264 & 219	& 156 & 134	& 122 & 94 & 4057 & 3767 & 290 \\
		\hline
	\end{tabular}
	\caption{The quantitative summary of the S1322 and PDBbind v2016 data sets.}
	\label{table:bd-set}
\end{table}

\paragraph{}
The ligand-based model is used in the present work. For the S1322 set, a 5-fold cross validation was conducted with the GBDT method. To be consistent with the results in the previous literature, accuracy is in term of Pearson correlation coefficient ($R$). Because the results from Daylight and Pharm2D fingerprints are relatively poor, their results are omitted here. The performance of the other six fingerprints (ECFP, FP2, Estate2, MACCS, Estate1, ERG) and their consensus are shown in Figure ~\ref{fig:s1322-bd}.

\paragraph{}
Figure ~\ref{fig:s1322-bd} indicates that for all the seven clusters, the consensuses of the six fingerprints largely achieve better performance than that of any single fingerprint. Specifically, the $R$ values of  consensus models are 0.717, 0.847, 0.708, 0.718, 0.831, 0.777, and  0.760 on each of 7 clusters, respectively and 0.765 on average. These results are comparable to ones achieved by a ligand-based 3D topology and GBDT model \cite{cang2018integration}.

\subsubsection{PDBbind v2016 refined set and core set}

\begin{figure}[ht!]
	\includegraphics[width=0.6\textwidth]{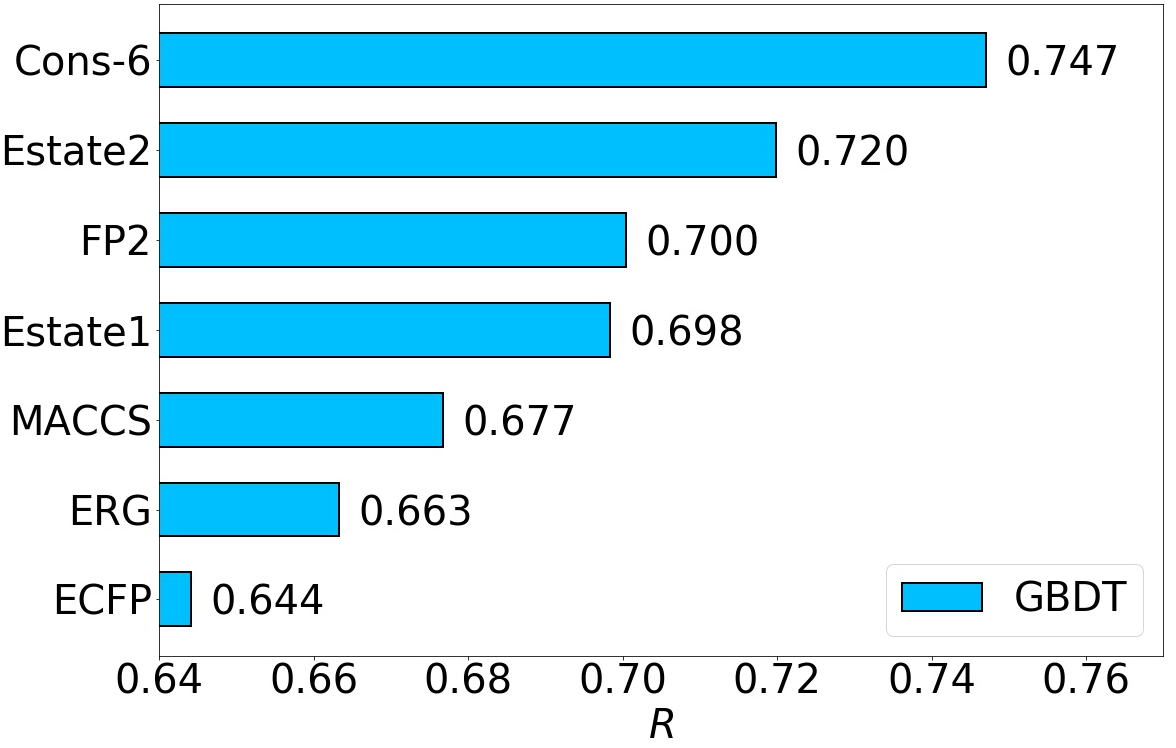}
	\caption{The $R$ on the PDBbind v2016 binding affinity set yielded by the six fingerprints (ECFP, FP2, Estate 2, MACCS, Estate 1, ERG) and their consensus.}
	\label{fig:bd2016-result}
\end{figure}
\paragraph{}
The present ligand-based model was also tested on PDBbind v2016. Rather than cross validation, this time the core set is regarded as a test set. Quite consistent with core validation on the S1322 set, the consensus of the six fingerprints leads to a large improvement than any single one, with an $R$ of 0.747.
These results indicate that the present model has a stable and reliable performance on different
protein-ligand binding affinity data sets.

\begin{table}[h!]
	\centering
	\begin{tabular}{|ccc|}
		\hline
		\rowcolor{Lightgreen}
		\textbf{Method} & \textbf{$R$}  & \textbf{RMSE (kcal/mol)} \\
		\hline
		 TopBP (Complex) \cite{cang2018representability} & 0.861 & 1.65 \\
			PLEC FP (Complex) \cite{wojcikowski2018development} & 0.817 & 1.71  \\
			Our cons-top 6  (Ligand) & 0.747 & 2.02\\
		\hline
	\end{tabular}
	\caption{Comparison of  protein-ligand binding affinity predictions PDBbind v2016 core set.}
	\label{table:pdbbindv2016-compare}
\end{table}

\paragraph{}
For protein-ligand binding affinity prediction, the present 2D fingerprint-based model is not competitive, because protein-ligand binding not only depends on the ligand, but also on the protein. Therefore, for a more accurate prediction, the information of the protein, at least the information of the binding site should be included. State differently, a complex based model is recommended. 
Recently,  Wojcikowski {\it et. al.}  \cite{wojcikowski2018development} reports 2D fingerprint-based  
complex models.  In their work, a recently developed 2D fingerprint model  is used to encode protein-ligand complex information. When combined with DNN, their method  gives rise to  an $R$ of 0.817 on the PDBBind v2016 core set. A complicated 3D structure-based model using the topology of the protein-ligand complex developed by our group  \cite{cang2018representability} has an $R$ of 0.861 on the same set. Table \ref{table:pdbbindv2016-compare} lists these results. 

\section{Discussion}

\subsection{General analysis  } 
In the present work, the predictive power of eight popular 2D fingerprints as well as their consensuses on four important drug-related properties (i.e., toxicity, Log S, Log P, binding affinity) was  investigated. The present study reveals that with a proper machine learning algorithm, the 2D fingerprint-based models including their consensuses outperform other 2D QSPR approaches in the most cases, especially on the toxicity predictions. Additionally, 2D fingerprint-based models  are comparable to state-of-the-art 3D structure-based models in most drug-related property predictions, except for protein-ligand binding affinity prediction. Considering 2D fingerprints are very "cheap" molecular descriptors that are easy and fast to generate, our results are very impressive. It means that 2D fingerprints  with appropriate machine learning algorithms are still very valuable for practical problems, such as the prediction of toxicity, the aqueous solubility (Log S), and the partition coefficient (Log P). However,  for protein-ligand binding affinity prediction, complex-based models using 3D topological fingerprints have a major advantage  over the present 2D fingerprints, i.e., about 15\% more accurate \cite{cang2018integration}. 

\subsection{The performance analysis of  2D fingerprints}
\subsubsection{Analysis of   2D fingerprints for  PDBbind v2016 core set predictions} 
\paragraph{}

\begin{figure}[ht!]
	\includegraphics[width=0.6\textwidth]{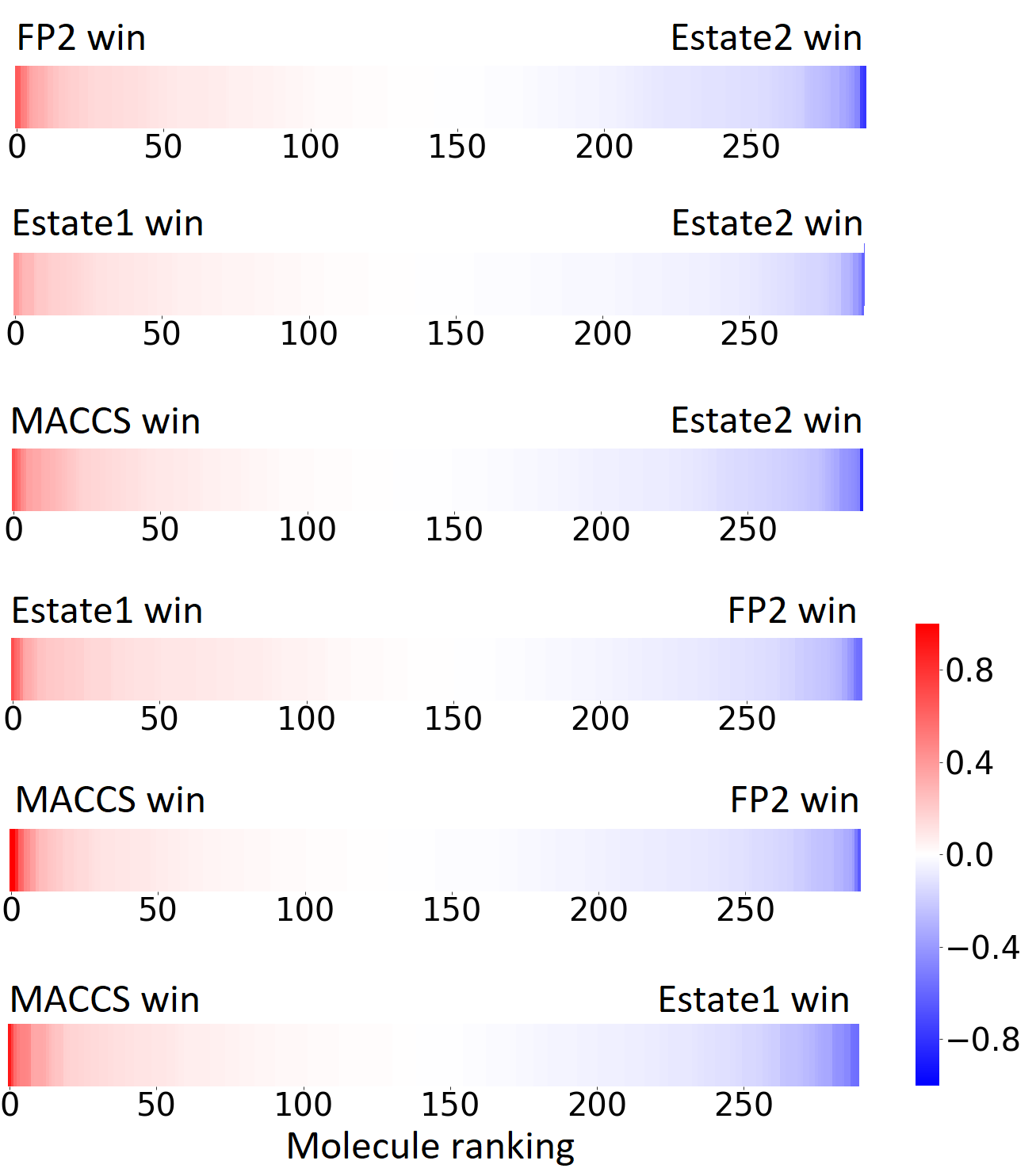}
	\caption{The ranked error differences between pairs of fingerprints for PDBbind v2016 core set of 290 molecules. Only the top 4 fingerprints    
 (i.e., Estate2, FP2, Estate1, MACCS) are considered. 
}
	\label{fig:beatmap-bd2016}
\end{figure}

The performance of each 2D fingerprint can be systematically analyzed by comparing the difference between prediction errors of every pair of fingerprints as follows.
\paragraph{}
(1) The relative absolute error for the $f$th fingerprint on the $i$th sample (molecule) in the test set  is defined by  
\begin{equation}\nonumber
 	{\rm Error}_{f,i} = \frac{| {\rm prediction ~ value}_{f,i}- {\rm experimental~ value}_i|}{| {\rm experimental~ value}_i|} 
\end{equation}

\paragraph{}
(2) For each molecule, the error difference between each pair of fingerprints is calculated. 

\paragraph{}
(3) Then, the differences for all molecules are ranked from the largest to smallest.  The result for PDBbind v2016 core set of 290 complexes is plot
in  Figure ~\ref{fig:beatmap-bd2016}. We have shown all of 6 pairs for the top four 2D fingerprints.  

\paragraph{} 
(4) To further analyze the strength of each fingerprint on certain molecules, we collect those molecules on which a fingerprint is able to outperform another fingerprint by 0.4 in the error difference. 

\paragraph{}
(5) Among these molecules for each fingerprint, we identify the top 10 most frequently occurred functional groups. The frequency of the occurrence of each functional group, along with the total of number of molecules, are given in Table ~\ref{table:fuc-group_bd2016}.


\LTcapwidth=\linewidth
\begin{longtable}{|c|P{3.5cm}|P{3.5cm}|P{3.5cm}|P{3.5cm}|}
	\hline
	\rowcolor{Lightgreen}
	\textbf{Ranking} & \textbf{FP2} & \textbf{Estate1} & \textbf{Estate2} & \textbf{MACCS} \\
	\hline
	& & & & \\
	1 & \includegraphics[scale=0.8]{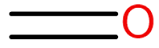} \newline carbonyl group: 24/41 & \includegraphics[scale=0.8]{carbonyl-group.png} \newline carbonyl group: 25/42 & \includegraphics[scale=0.8]{carbonyl-group.png} \newline carbonyl group: 23/41 & \includegraphics[scale=0.8]{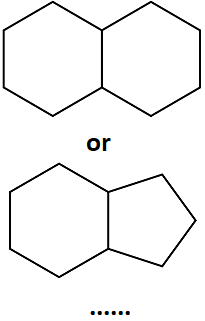} \newline bicyclic compounds: 17/36 \\
	\hline
	& & & & \\
	2 & \includegraphics[scale=0.8]{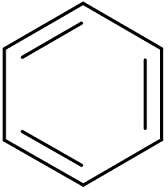} \newline unfused benzene ring: 21/41 & \includegraphics[scale=0.8]{benzene.png} \newline unfused benzene ring: 18/42 & \includegraphics[scale=0.8]{benzene.png} \newline unfused benzene ring: 22/41 & \includegraphics[scale=0.8]{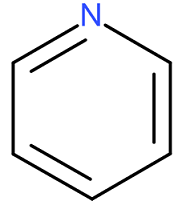} \newline pyridine: 17/36 \\
	\hline
	& & & & \\
	3 & \includegraphics[scale=0.8]{bicyclic.png} \newline bicyclic compounds: 19/41 & \includegraphics[scale=0.8]{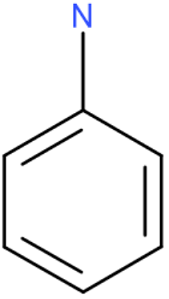} \newline aniline:14/42 & \includegraphics[scale=0.8]{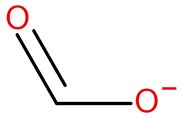} \newline carboxylate ion: 16/41 & \includegraphics[scale=0.8]{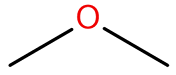} \newline ether: 16/36 \\
	\hline
	& & & & \\
	4 & \includegraphics[scale=0.8]{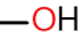} \newline hydroxyl: 16/41 & \includegraphics[scale=0.8]{carboxylate.png} \newline carboxylate ion: 14/42 & \includegraphics[scale=0.8]{bicyclic.png} \newline bicyclic compounds: 15/41 & \includegraphics[scale=0.8]{carbonyl-group.png} \newline carbonyl group 15/36 \\
	\hline
	& & & & \\
	5 & \includegraphics[scale=0.8]{ether.png} \newline ether: 14/41 & \includegraphics[scale=0.8]{oh.png} \newline hydroxyl: 14/42 & \includegraphics[scale=0.8]{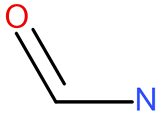} \newline carbonyl group with N: 13/41 & \includegraphics[scale=0.8]{oh.png} \newline hydroxyl: 15/36 \\
	\hline
	& & & & \\
	6 & \includegraphics[scale=0.8]{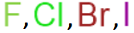} \newline 12/41 & \includegraphics[scale=0.8]{ether.png} \newline ether: 14/42 & \includegraphics[scale=0.8]{ether.png} \newline ether: 13/41 & \includegraphics[scale=0.8]{benzene.png} \newline unfused benzene ring: 12/36 \\
	\hline
	& & & & \\
	7 & \includegraphics[scale=0.8]{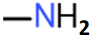} \newline amide: 10/41 & \includegraphics[scale=0.8]{carboxylate-n.png} \newline carbonyl with Nitrogen: 13/42 & \includegraphics[scale=0.8]{oh.png} \newline hydroxyl: 11/41 & \includegraphics[scale=0.8]{nh2.png} \newline amide: 10/36 \\
	\hline
	& & & & \\
	8 & \includegraphics[scale=0.6]{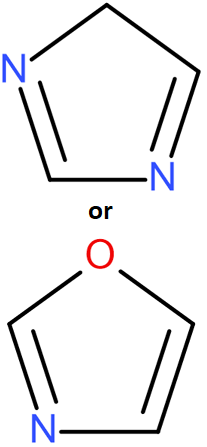} \newline azole: 8/41 & \includegraphics[scale=0.8]{nh2.png} \newline amide: 11/42 & \includegraphics[scale=0.8]{nh2.png} \newline amide: 11/41 & \includegraphics[scale=0.8]{carboxylate.png} \newline carboxylate ion: 9/36 \\
	\hline
	& & & & \\
	9 & \includegraphics[scale=0.6]{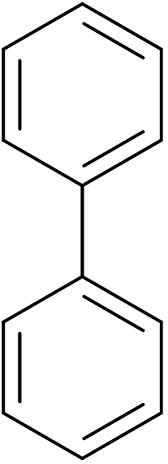}...... \newline multiple non-fused benzene rings: 7/41 & \includegraphics[scale=0.8]{fclbri.png} \newline 11/41 & \includegraphics[scale=0.8]{aniline.png} \newline aniline: 10/41 & \includegraphics[scale=0.6]{azole.png} \newline azole: 7/36 \\
	\hline
	& & & & \\
	10 & \includegraphics[scale=0.8]{aniline.png}...... \newline aniline: 7/41 & \includegraphics[scale=0.8]{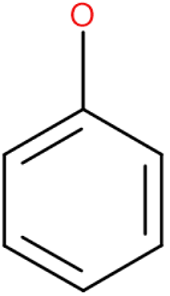} \newline phenol: 8/42 & \includegraphics[scale=0.6]{multi-ring.png}...... \newline multiple non-fused benzene rings 8/41 & \includegraphics[scale=0.8]{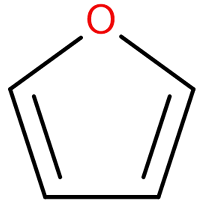} \newline furan: 5/36 \\
	\hline
	\caption{The top 10 frequently occurred functional groups in PDBbind v2016 core set for each fingerprint.  For each fingerprint, the occurrence  frequency and 	 the total number of molecules are also given. 
	 }
	\label{table:fuc-group_bd2016}
\end{longtable}

\paragraph{}
This  analysis is quite significant as shown in Table ~\ref{table:fuc-group_bd2016}. It indicates that different fingerprints have different performance on certain functional groups: some fingerprints perform better on some functional groups, while other fingerprints perform better on other functional groups. Therefore, one can select an appropriate fingerprint to represent a certain class of functional groups based on Table ~\ref{table:fuc-group_bd2016}.  For the FP2, Estate1, and Estate2 fingerprints, the top two  functional groups are   carbonyl groups and unfused benzene rings. However, the MACCS fingerprint is some special. Its top two functional groups  are bicyclic compounds and pyridine. 



The third top functional groups differ much  for   four fingerprints: bicyclic compounds for FP2, aniline for Estate1, carboxylate ion for Estate2, and ether for MACCS, which gives us more information to choose fingerprints. Such as, if one has a molecule including aniline, then Estate1 should be selected. Noticeably, some functional groups occur  exclusively for one or two types of fingerprints. For example,  F, Cl, Br, I is only on the lists of FP2 and Estate1. While azole appears only on the list of  FP2 and MACCS and multiple non-fused benzene rings are only for FP2 and Estate 2. Moreover, phenol occurs only for Estate1 and furan occurs only for MACCS.

\subsubsection{Analysis of 2D fingerprints for the IGC$_{50}$ toxicity data set prediction and also other data sets} 

\begin{figure}[ht!]
	\includegraphics[width=0.6\textwidth]{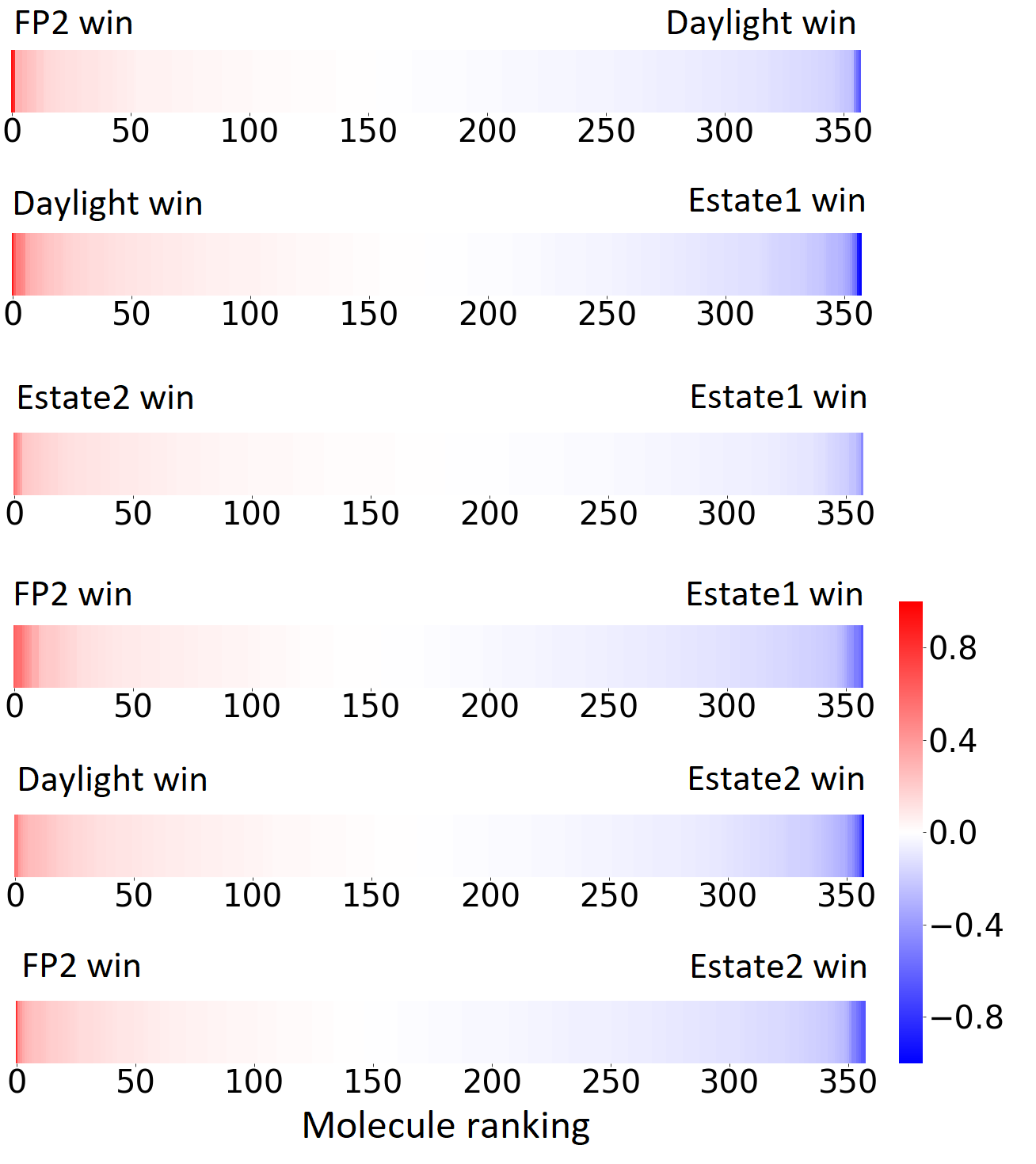}
	\caption{The ranked error differences between pairs of fingerprints for IGC$_{50}$ toxicity set of 358 molecules. Only the top 4 fingerprints     (i.e.,  Estate2, FP2, Estate1, Daylight) are considered. 
	}
	\label{fig:beatmap-tox}
\end{figure}

Using the same 5-step procedure outlined above, we carry out a performance analysis for toxicity dataset IGC$_{50}$, which is shown in Figure ~\ref{fig:beatmap-tox} and Table ~\ref{table:fuc-group_tox}. The molecules in the toxicity data set are typically small and simple, leading to the functional groups in Table ~\ref{table:fuc-group_tox} also small.  Moreover, since there are not too many functional groups in these relatively simple molecules, only top 8 functional groups are presented in the table. Similar to the performance on the binding affinity, for the top 4 fingerprints on the toxicity set, the carbonyl group is in the first place. Unfused benzene rings also have a high occurrence frequency, resulting in the second or third ranking. The difference between the performance of various fingerprints is mainly located on sulfide and aliphatic chains with  8 or more members. FP2 fingerprint works well on sulfide, whereas, Daylight, Estate1 and Estate2  work well on aliphatic chains with 8 or more members.

\paragraph{}

\begin{longtable}{|c|P{3.5cm}|P{3.5cm}|P{3.5cm}|P{3.5cm}|}
	\hline
	\rowcolor{Lightgreen}
	\textbf{Ranking} & \textbf{FP2} & \textbf{Daylight} & \textbf{Estate1} & \textbf{Estate2}  \\
	\hline
	& & & & \\
	1 & \includegraphics[scale=0.8]{carbonyl-group.png} \newline carbonyl group: 14/33 & \includegraphics[scale=0.8]{carbonyl-group.png} \newline carbonyl group: 17/34 & \includegraphics[scale=0.8]{carbonyl-group.png} \newline carbonyl group: 16/39 & \includegraphics[scale=0.8]{carbonyl-group.png} \newline carbonyl group: 14/37 \\
	\hline
	& & & & \\
	2 & \includegraphics[scale=0.8]{benzene.png} \newline unfused benzene ring: 21/41 & \includegraphics[scale=0.8]{oh.png} \newline hydroxyl: 9/34 & \includegraphics[scale=0.8]{oh.png} \newline hydroxyl: 15/39 & \includegraphics[scale=0.8]{oh.png} \newline hydroxyl: 14/37 \\
	\hline
	& & & & \\
	3 & \includegraphics[scale=0.8]{nh2.png} \newline amide: 9/33 & \includegraphics[scale=0.8]{benzene.png} \newline unfused benzene ring: 9/34 & \includegraphics[scale=0.8]{benzene.png} \newline unfused benzene ring: 9/39 & \includegraphics[scale=0.8]{benzene.png} \newline unfused benzene ring: 10/37 \\
	\hline
	& & & & \\
	4 & \includegraphics[scale=0.8]{oh.png} \newline hydroxyl: 9/33 & \includegraphics[scale=0.8]{nh2.png} \newline amide: 7/34 & \includegraphics[scale=0.8]{ether.png} \newline ether: 7/39 & \includegraphics[scale=0.8]{ether.png} \newline ether: 8/37 \\
	\hline
	& & & & \\
	5 & \includegraphics[scale=0.8]{ether.png} \newline ether: 8/33 & \includegraphics[scale=0.8]{ether.png} \newline ether: 7/34 & \includegraphics[scale=0.8]{fclbri.png} \newline 6/39 & \includegraphics[scale=0.8]{fclbri.png} \newline 8/37 \\
	\hline
	& & & & \\
	6 & \includegraphics[scale=0.8]{fclbri.png} \newline 5/33 & \includegraphics[scale=0.8]{fclbri.png} \newline 6/34 & \includegraphics[scale=0.8]{nh2.png} \newline amine: 6/39 & \includegraphics[scale=0.8]{nh2.png} \newline amine: 6/37 \\
	\hline
	& & & & \\
	7 & \includegraphics[scale=0.8]{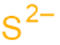} \newline sulfide: 3/33 & aliphatic chains with 8 or more members: 5/34 & aliphatic chains with 8 or more members: 6/39 & aliphatic chains with  8 or more members: 5/37 \\
	\hline
	& & & & \\
	8 & \includegraphics[scale=0.8]{aniline.png} \newline aniline: 3/33 & \includegraphics[scale=0.8]{aniline.png} \newline aniline: 4/34 & \includegraphics[scale=0.8]{aniline.png} \newline aniline: 3/39 & \includegraphics[scale=0.8]{aniline.png} \newline aniline: 5/37 \\
	\hline
	\caption{The top 10 frequently occurred functional groups in  IGC$_{50}$ toxicity set  for each fingerprint.  For each fingerprint, the occurrence  frequency and 	 the total number of molecules are also given. 
 }
	\label{table:fuc-group_tox}
\end{longtable}

\paragraph{}
The same performance analyses were also conducted for other toxicity and log P data sets, the results are shown in Tables S1 to S4. These tables indicate, for the toxicity data sets of LD$_{50}$, LC$_{50}$, LC$_{50}$-DM, the performance of the Estate1 and Estate2 fingerprints are similar, they both work well on bicycle compounds; comparing to it, the FP2 fingerprint works better on aliphatic chains with 8 or more members, the daylight fingerprint has a better performance on amide. For log P data set, the ECFP and Estate2 fingerprints lead to a good performance on aniline, the Estate 1 fingerprints works better on bicycle compounds; MACCS fingerprint works better on unfused benzene ring.  

\subsection{The predictive power of the consensus of 2D fingerprints}

\paragraph{}
The consensus of several different fingerprints typically further enhances the performance of a single fingerprint. This enhancement can be  quite significant. However, on the datasets of different drug-related properties, the best fingerprint combinations for the consensus are not consistent. One possible explanation is that different fingerprints are good at encoding certain functional groups, and datasets for different drug-related properties have different functional group distributions.  This is also the reason why a consensus can enhance performance. The consensus can capture more functional groups and counter-balance the systematical bias from different fingerprints.

\paragraph{}
On toxicity prediction, the best combination for consensus is obtained with Estate2, Estate1, Daylight, FP2, ECFP, and MACCS. On the Log S prediction, the best combination is achieved with MACCS, Estate1, and Daylight. While on the Log P prediction, the best  consensus involves Estate2, Estate1, ECFP, and MACCS.  Finally, on the binding affinity prediction, the best consensus uses Estate2, Estate1, FP2, ECFP, MACCS, and ERG. It is worth noting that, Estate related (Estate1, Estate2 or both) models are always included in the best combinations. In fact,  their single performances are relatively good. This finding is not surprising since  Estate fingerprints encode the intrinsic electronic state of the atom as perturbed by the electronic influence of all other atoms. It is well-known that electronic state is important to drug-related properties.

\subsection{Multitask deep learning}
\paragraph{}
Multitask deep learning was utilized on our toxicity prediction. It turns out that the smallest set LC$_{50}$-DM with only 283 training samples benefits dramatically from the multitask deep learning strategy. Its  $R^2$ value rises from 0.523 to 0.725. This is because, in the frame of multitask deep learning, different data sets (tasks) share similar structure-function relationships.  When a small dataset is trained with a large dataset through shared neural networks, the statistics learned from the large datasets in the shared neurons can help predict the small dataset property.  As a result, the other three large toxicity sets can share their patterns  learned from training with the small toxicity set, enhancing its prediction. Therefore,  multitask deep learning could be a useful strategy to train relatively small datasets.

\subsection{The limitation and advantage of 2D fingerprints}
\paragraph{}
Typically, 2D fingerprints only encode small molecules, such as  ligands, although high level 2D fingerprint models including both proteins and ligands have also been developed \cite{wojcikowski2018development,kundu2018machine}. Theoretically, 2D fingerprints are more suitable for target-independent or target-unspecific problems involving small molecules, such as toxicity, solvation free energy, aqueous solubility,  partition coefficient, permeability, etc.  The current investigation confirms this point.  For toxicity, aqueous solubility and partition coefficient, the present 2D-fingerprint based methods perform quite similar to or even somewhat better than 3D structure-based methods  in some cases. 

For protein-ligand binding affinity predictions, both  ligand-based approaches and complex-based are examined.  For ligand-based approaches,  2D-fingerprint based methods  can perform as  well as 3D structure-based models.  However, 3D structure-based topological models \cite{cang2018representability} outperform   2D-fingerprint based methods (i.e., R: 0.861 vs 0.747 for PDBbind v2016 core test). In fact,  more sophisticated 2D fingerprint models that utilize the protein-ligand complex information and DNN \cite{wojcikowski2018development,kundu2018machine} are still not as accurate as 3D topology-based models \cite{cang2018representability}(i.e., R: 0.817 vs 0.861 for PDBbind v2016 core test and  0.774 vs 0.808 for PDBbind v2013 core test).  Essentially, algebraic topology is designed to simplify the geometric complexity of biological macromolecules. Therefore, it is able to extract vital information from protein-ligand complexes to predict their binding affinities. 

2D fingerprints are much easier to generate than 3D structure-based fingerprints built form algebraic topology, differential geometry or various graph theory. Therefore,  2D-fingerprint based models can be useful tools for preliminary drug screening studies. 

\section{Conclusion}
Two-dimensional molecular fingerprints, or 2D fingerprints,  refer to molecular structural patterns, such as elemental composition, atomic connectivity, functional groups, 2D-pharmacophores etc. extracted from a molecule without taking into account the 3D-structural representation of these properties. 2D fingerprints have been a main workhorse  for cheminformatics and bioformatics for decades.  However,  their validations in various datasets  were typically carried out long time ago with earlier machine learning algorithms.  Recently,  new 3D structure-based molecular fingerprints built from algebraic topology \cite{cang2018integration,cang2018representability}, differential geometry \cite{nguyen2019dg}, geometric graph theory \cite{nguyen2017rigidity, bramer2018multiscale}, and algebraic graph theory \cite{nguyen2019mathematical} have found much success in drug discovery  related applications \cite{wu2018quantitative,wu2018topp,cang2018integration,cang2018representability}, including D3R Grand Challenges \cite{gaieb2019d3r,nguyen2019mathematical}. It raises an interesting issue whether 2D fingerprints are still competitive in drug discovery related applications. 

This work reassesses 2D fingerprints for their performance in drug discovery related applications. We consider a total of eight commonly used 2D fingerprints, namely FP2, Daylight, MACCS, Estate1, Estate2,  ECFP,  Pharm2D,  and ERG.
Four types of drug discovery related applications with 23 datasets,  including solubility (Log S) and partition coefficient (Log P) that are independent of a target protein, toxicity that may depend on certain unknown target proteins, and protein-ligand binding affinity that depend on known target proteins,  are designed to validate 2D  fingerprints.  Advanced machine learning algorithms, including random forest (RF), gradient boosting decision trees (GBDT), single-task deep neural network (ST-DNN), and  multitask deep neural network (MT-DNN) are used to optimize the performance of the above 2D fingerprints in the aforementioned four types of datasets. In particular, MT-DNN is designed to enhance the performance of 2D fingerprints on relatively small datasets by a simultaneous training with relatively large datasets that share a similar pattern.   Since each fingerprint may have an explicit bias on certain functional groups or 2D patterns, we carry out various consensus to further boost the performance of 2D fingerprints in all the datasets. Finally, the strengths of top four 2D fingerprints for predicting protein-ligand binding  affinity and quantitative toxicity are analyzed in detail.   

Our general findings are as follows. 
1) 2D fingerprint-based models are as good as 3D structure-based models for various toxicity, Log S and Log P datasets under the same training-test condition.  
2) For ligand-based protein-ligand binding affinity predictions, 2D fingerprint-based models perform equally well as 3D structure-based models that are based only on ligand 3D structures.   
3) 3D structure-based models that utilize 3D protein-ligand complex information outperform 2D fingerprints that based on either ligand information or protein-ligand complex information.    
4) Advanced machine learning algorithms, such as DNN and MT-DNN, are crucial for 2D  fingerprints to achieve optimal performance. 
5) There is no 2D fingerprint that outperforms all other 2D fingerprints in all applications. 
6) Appropriate consensus of a few 2D models typically achieves better performance. 
Therefore, if combined with advanced machine learning algorithms, the 2D fingerprints are still competitive in most drug discovery related applications except for those that involve protein structures.

 \section*{Acknowledgments}
This work was supported in part by  NSF Grants DMS-1721024,  DMS-1761320, and IIS1900473 and NIH grant  GM126189.  

\clearpage 


%
%

\end{document}